\documentclass[12pt]{article}
\usepackage{epsf}
\usepackage{graphicx}
\usepackage{amssymb}
\usepackage{epstopdf}
\usepackage{amsmath}
\usepackage{amssymb}
\usepackage{epsfig}
\usepackage{cite}
\usepackage{color,colordvi}
\usepackage{graphicx}
\usepackage{pdfpages}
\graphicspath{ {./} }
\usepackage[left=2.5cm,bottom=3cm,right=2.5cm,top=3cm]{geometry} 

\usepackage[latin,english]{babel}
\usepackage{amsmath}
\usepackage{amssymb}
\usepackage{graphicx}
\usepackage{mathtools}
\usepackage{tikz} 
\usetikzlibrary{patterns}
\usetikzlibrary{decorations.markings,decorations.pathmorphing}
\usetikzlibrary{calc}
\tikzstyle{singularity}=[line width=0.6,decorate,
                         decoration={zigzag,amplitude=2,segment length=6.17}]
\usetikzlibrary{shapes.misc}
\colorlet{mydarkred}{red!50!black}
\begin{document}
\vspace{0.01cm}
\begin{center}
{\Large\bf  The  Algebraic Page Curve} 

\end{center}

\vspace{0.1cm}

\begin{center}

{\bf Cesar Gomez}$^{a}$\footnote{cesar.gomez@uam.es}

\vspace{.6truecm}

{\em $^a$
Instituto de F\'{\i}sica Te\'orica UAM-CSIC\\
Universidad Aut\'onoma de Madrid,
Cantoblanco, 28049 Madrid, Spain}\\

\end{center}

\begin{abstract}
\noindent  
 
{The Page curve describing the process of black hole evaporation is derived in terms of a family, parametrized in terms of the evaporation time, of finite type $II_{1}$ factors, associated, respectively, to the entanglement wedges of the black hole and the radiation. The so defined Page curve measures the relative continuous dimension of the black hole and the radiation along the evaporation process. The transfer of information is quantitatively defined in terms of the Murray von Neumann parameter describing the change of the spatial properties of the factors during the evaporation. In the simplest case the generator of the evaporation process is defined in terms of the action of the fundamental group of the hyperfinite type $II_1$ factor. In this setup the Page curve describes a phase transition with the transfer of information as order parameter. We discuss the limits of either a type $I$ 
or a type $III$ description of the black hole evaporation.
}

\end{abstract}

\thispagestyle{empty}
\clearpage
\tableofcontents
\newpage

\section{Introduction}
Hawking's radiation \cite{Hawking1} is a well understood consequence of the semiclassical description of quantum field theory in the curved background defined by the black hole classical space-time geometry. At this semiclassical level the Hawking radiation is, as view by the external observer, thermal. Its quantum mechanical description through a reduced density matrix leads us to quantify the entanglement between the radiated quanta and the black hole interior as the von Neumann entropy of this density matrix. Semiclassically this vN entropy grows with time.

If we extend the limits of applicability of this semiclassical analysis, up to scales where we expect quantum gravity effects will dominate, we get the semiclassical picture of black hole evaporation. 

Hawking {\it information paradox} \cite{Hawking2} appears, in this semiclassical context ,at the moment in which the Bekenstein Hawking thermodynamic black hole entropy \cite{Bek1}, becomes smaller than the von Neumann entanglement entropy of the radiation. Paradoxically this can happen at scales semiclassically very safe where the black hole size is much larger than any quantum gravity scale.

A seminal observation of Page \cite{Page} is that if we {\it model}, at any time of the evaporation process, the black hole and the radiation as defining a bipartite quantum mechanical system in a Hilbert space of {\it finite dimension} $N$, then it automatically follows that the maximal entanglement entropy of the radiated quanta is reached after $O(\log\sqrt{N})$ quanta are emitted. Moreover any model of the evaporation process consistent with a bipartite quantum mechanical description in a finite dimensional Hilbert space defines a Page like curve.

This observation leads to the introduction of the so called {\it Central Dogma} \cite{Malda} that imposes that, from the point of view of the outside, the black hole evaporation can be described using a {\it finite dimensional} Hilbert space. We do not need to insist on how revolutionary this dogma may seem at first glance. The usual justification of the dogma is based on identifying the dimension of the Hilbert space in terms of the finite BH thermodynamic entropy $S_{BH}$ as $e^{S_{BH}}$. In any case note that once the dogma about the finitude of the Hilbert space is accepted the Page curve will appear by default. In this note we will suggest a way to derive the Page curve using an infinite dimensional Hilbert space of quantum states.

But before discussing how to derive the Page curve using an infinite dimensional Hilbert space of quantum states we should address a preliminary question, namely; Can we identify an improved semiclassical definition of the radiation entanglement entropy that, even semiclassically, reproduces the expected behaviour predicted by Page ? If the answer is yes then we need to figure out how this semiclassical Page curve is related to the finitude of the underlying Hilbert space.

In more general terms Page's curve follows from the {\it unitarity} of the black hole evaporation process. This unitarity is pretty natural if we can map the gravitational evaporation process into a unitary evolution in a QFT dual ( through some AdS/CFT correspondence). Thus a natural question is how the finitude of the Hilbert space, assumed in the dogma, is encoded in the QFT dual description of the evaporation process.

In recent years and motivated by the holographic RT \cite{RT}\cite{HRT},\cite{FLM} representation of entanglement entropy, a beautiful semiclassical definition of the black hole and the radiation entropy, consistent with Page curve and consequently with unitarity, has been developed using the QES prescription \cite{EW},\cite{EW1},\cite{EW2},\cite{EW3}\cite{EW4}.

The crucial aspect of the QES prescription consists of identifying a generalized entropy definition \cite{Bek2},\cite{Bek3} ( see also \cite{susskind1}) of the entanglement entropy of the radiated quanta as well as a generalized entropy of the black hole entropy \footnote{For other discussions of Page curve and mutual information see \cite{mutual1}\cite{mutual2}}. 

But, What is the meaning of these {\it generalized entropies}?
It is in order to answer this question that the recent algebraic approach to gravity becomes crucial (for a sampling of papers used in the rest of this note see \cite{Liu1,gomez2}).

Generically entanglement entropy is defined as the von Neumann entropy of some reduced density matrix accounting for the full information available to an observer that can perform experiments in a bounded region of space time. In the case of an eternal black hole we can identify, for the external observer, the black hole horizon as the boundary of this region. Algebraically we can associate a von Neumann algebra $M$ ( or even a factor ) to this region. If this factor admits a {\it trace state} i.e. a linear form $\tau_M$ such that $\tau_M(ab)=\tau_M(ba)$ for $a$,$b$ arbitrary elements of $M$, then we can associate with any state $\phi$ on $M$ i.e. a linear form on $M$ that assigns to any observable $a\in M$ its expectation value $\phi(a)$, a {\it density matrix} $\rho_{\phi}$. Incidentally this density matrix is a self adjoint operator, as it should be the case for any density matrix, that generically is not in $M$ but affiliated to $M$ \footnote{The notion of affiliation goes back to the work \cite{MvN1}.}

A very interesting result in \cite{Witten2} is that the von Neumann entropy of these affiliated density matrix \cite{Longo} can be interpreted, for the case of the AdS eternal black hole, as generalized entropies in the sense used in the QES prescription. Note that this result follows from the general relation $\rho_{\phi}^{it}=u_{\phi,\tau_M}(t)$ \cite{CT} defining the affiliated density matrix from Connes cocycle \cite{Connes} between the state $\phi$ and the tracial state $\tau_M$ \footnote{ From a physics point of view the interpretation of the von Neumann entropy of these affiliated density matrix as generalized entropies becomes clear when we use, in order to define the trace $\tau_M$, a dominant weight \cite{CT}. This will be discussed in section 4.1.}.

But, How can we extend this result to the black hole evaporation process ? or in other words, How can we define {\it algebraically} a time dependent generalized entropy consistent with Page curve? This will be the question we will address in this note.
 
 \vspace{0.3 cm}
 
When describing the process of black hole evaporation we need to deal, at each evaporation time, with two factors. One factor associated with the black hole entanglement wedge and the other associated with the radiation entanglement wedge. We will generically denote these two factors as $M_B(t)$ and $M_R(t)$. What is the relation between these two factors? 

\vspace{0.3 cm}

Our {\it first assumption} in this note will be that, at any evaporation time $t$, the factors satisfy the commutant relation $M_B(t)^{'}=M_R(t)$. Note that this assumption implies, besides the algebraic properties of both factors, the Hilbert space on which they are acting. Recall that the commutant of a factor is defined relative to the Hilbert space on which the factor is acting. 

At this point is important to remember the basic distinction introduced in \cite{MvN4} between {\it algebraic} and {\it spatial} properties. While {\it algebraic properties} are preserved by any isomorphism that preserves the algebraic defining properties of the algebra as for instance the multiplication rule and the identity, {\it spatial properties} depend on the particular Hilbert space $H$ on which the factor is acting. Thus the notion of commutant is a spatial property. In what follows whenever we refer to two factors $M_1$ and $M_2$ such that $M_1^{'}=M_2$ we will denote these factors as a couple $(M_1,M_2)$. In this sense our {\it first assumption} is that the factors associated to the black hole and radiation entanglement wedges define a couple $(M_B(t),M_R(t))$ and that this is the case at any evaporation time $t$.  

\vspace{0.3 cm}

Our {\it second assumption} is that {\it the evaporation process will be characterized by the time evolution of the {\bf spatial properties} of the couple $(M_B(t),M_R(t))$}. In other words the evaporation process depends on the interplay between $M_B(t)$ and $M_R(t)$ and not only on the algebraic properties of $M_B(t)$ and $M_R(t)$ at different times. It is this interplay between the black hole and the emitted radiation what we will describe using {\it spatial} properties. In that sense an eternal black hole will be generically associated with a family of couples $(M_B(t),M_R(t))$ that are spatially isomorphic among them. 

\vspace{0.3 cm}

Finally our {\it third assumption} will be that the factors $M_B(t)$ and $M_R(t)$ will be, at any $t$, {\it finite factors}.

\vspace{0.3 cm}

Note that the {\it Central Dogma} that assumes the finitude of the Hilbert space on which the factors are acting satisfies the three assumptions. Indeed in such a case the factors $M_B(t)$ and $M_R(t)$ are {\it finite type $I$ factors} and a Page like curve is automatic for the evolution of any spatial property of this couple of factors. The simplest possibility is to use as spatial property the relative Murray von Neumann dimensions that in this simple case are fully determined by the finite dimensions of the corresponding factorization of the finite dimensional Hilbert space.  

However this type $I$ model of our assumptions neither explains the appearance of generalized entropies and the associated QES's nor the mechanism of information transfer taking place during the evaporation process.

\vspace{0.3 cm}

In this note we will relax the central dogma in a way consistent with our three assumptions. Namely {\it we will keep the finitude of the couple of factors $(M_B(t),M_R(t))$ but we will assume that they act on an {\bf infinite dimensional Hilbert space}}. This implies that these factors are necessarily finite type $II_1$ factors. Thus the black hole evaporation will be described in terms of

\vspace{0.3 cm}

 {\it The time dependence of the spatial properties of these couples of type $II_1$ factors.}

\vspace{0.3 cm}

The simplest and more basic spatial property we can use is the Murray von Neumann coupling that we will denote $d_{MvN}(B,t)$ for the black hole and {\it its dual} $d_{MvN}(R,t)=\frac{1}{d_{MvN}(R,t)}$ for the radiation. In these conditions the so defined {\it algebraic Page curve} will describe the time evolution during the evaporation process of 
 
 \vspace{0.3 cm}

{\it The relative continuous dimensions of the black hole and the radiation entanglement wedges}.

\vspace{0.3 cm}
 In this note we will present an explicit type $II_1$ model of black hole evaporation.
This type $II_1$ model produces several interesting outputs. First of all it identifies a quantitative {\it measure of the information transfer} along the evaporation process that is determined by the Murray-von Neumann parameter $\theta$ \cite{MvN4} directly determined by the couplings $d_{MvN}(B,t)$. Secondly it allows us to identify as generator of the evolution during the evaporation process the action, induced by Takesaki automorphisms \cite{Takesaki}, on {\it the fundamental group} of the hyperfinite type $II_1$  factor ( see \cite{Popa,Jones}).

\vspace{0.3 cm}

Along the paper we will discuss other algebraic representations \cite{EL2} of the black hole evaporation using couples of {\it infinite factors}. 

\vspace{0.3 cm}

In summary we find a type $II_1$ algebraic Page curve defining the time evolution, during evaporation, of the relative continuous dimensions of the black hole and the radiation entanglement wedges. In this picture the {\it Page time} defines a transition between two dual phases with the transfer of information parameter playing the role of the {\it order parameter}. Finally and inspired by a comment in \cite{Witten2}, on the potential algebraic meaning of $G_N$, we suggest an algebraic representative of $G_N$ in terms of the Murray von Neumann parameter $\theta$.

\section{Preliminary comments: the meaning of finiteness}
In \cite{MvN1} Murray and von Neumann addressed and solved the following problem. Given a Hilbert space $H$ classify all solutions of the equation
\begin{equation}\label{problem}
M\cap M^{'}={\mathbb{C}} 1
\end{equation}
where $M$ is a subalgebra of $B(H)$ and $M^{'}$ the corresponding commutant \footnote{The algebra $B(H)$ is the algebra of bounded operators acting on $H$. The commutant is defined as the algebra of operators in $B(H)$ commuting with all the elements in $M$.}. Solutions to (\ref{problem}) are factors. Murray and von Neumann identify two generic types of solutions: {\it finite} and {\it infinite}. The finite solutions are further classified as type $I_n$ and $II_1$ while the infinite solutions are characterized as $I_{\infty}$, $II_{\infty}$ and type $III$. A different classification of the solutions to (\ref{problem}) is to identify those defining {\it normal factors} i.e. those solutions $M,M^{'}$ for which it exists a decomposition of the Hilbert space $H=H_1\otimes H_2$ such that $M=B(H_1)$ and $M^{'}=B(H_2)$. The only solutions that are normal are those corresponding to $I_n$ and $I_{\infty}$. From a physics point of view and thinking $H$ as the Hilbert space of a given quantum mechanical system the solutions to (\ref{problem}) define the different decompositions into subsystems. The normal factors are the ones  corresponding to factorizations of the Hilbert space of states. 

The great physical insight of Murray and von Neumann was to push {\it finiteness} instead of {\it normality} and to highlight the special role of type $II_1$ that is finite but not normal by contrast to $I_{\infty}$ that is normal but infinite.

In {\it quantum information} we deal with $I_n$ with $n$ representing the number of q-bits. Pushing $n$ to infinite can give rise to the infinite factor $I_{\infty}$ but also, as it was first observed by von Neumann in \cite{vN}, using infinite tensor products of q-bits, to the hyperfinite type $II_1$ factor $R$. 

Hence, what is the meaning of finiteness and why is potentially more important, from a physics point of view, than normality? \footnote{The reader can find instructive the following quote from the introduction of \cite{MvN1}: {\it Which (factor)  has the most properties in
common with $I_n$, the ring of all matrices of $n$ rows and $n$ columns, and which
one can be considered as the limiting case of $I_n$ for $n =\infty$ ? The investigations
of operators in Hilbert space have always been carried on with the idea that the
(space of bounded operators) of Hilbert space, that is $I_{\infty}$, is the natural limiting case. We think however
that our results indicate that there is more point in assigning this role to type $II_1$}.}

By a finite factor we mean a factor $M$ such that for any hermitian operator $A\in M$ you can define a finite number playing the role of the {\it dimension} of the space of solutions in $H$ to the linear equation $Af=0$ i.e. the {\it rank} of $A$. Obviously this dimension is finite for the {\it quantum information} case $I_n$ but {\it it is also finite}, although not discrete, for finite factors of type $II_1$. The idea is that in such a case we can assign to each subspace $\Sigma$ of $H$, such that $P_{\Sigma}\in M$ for $P_{\Sigma}$ the corresponding projection, an appropriately normalized finite dimension $d_M(\Sigma)$ valued in $[0,1]$. More precisely, finiteness of a factor $M$ acting on $H$ means that it exists a linear map $d_M(P)$ on the space of projections in $M$ that assigns a finite {\it dimension} to the subspaces $PH$. This map $d_M$ can be defined axiomatically as a map from the space of projections into $\mathbb{R}^{+}$ satisfying the conditions i) $d_M(1)=1$, ii) $d_M(P)=d_M(Q)$ if $P$ and $Q$ are {\it equivalent projections} and iii) $d_M(P_1+P_2)=d_M(P_1)+d_M(P_2)$ if $P_1$ and $P_2$ are orthogonal.
For the infinite case $d_M$ satisfying the former conditions exists but it is not finite. Thus we can classify the solutions of (\ref{problem}) by the spectrum $Sp(d_M)$ of $d_M$. For the $I_n$ case this spectrum is the set $\{0,1,...n\}$, for type $II_1$ is the interval $[0,1]$. For the infinite cases we have $\{0,1....\infty\}$ for $I_{\infty}$, $[0,\infty]$ for type $II_{\infty}$ and $\{0,\infty\}$ for the type $III$ case. This classifies the full set of solutions to (\ref{problem}). 

For finite factors the dimension $d_M$ defines a {\it trace} or more properly a {\it tracial state} on $M$. As said, for $M$ a factor acting on the Hilbert space $H$ and $P$ a projection in $M$ the quantity $d_M(P)$ defines the {\it dimension} ( that can be continuous ) of the subspace $PH$, as the trace of $P$ \footnote{The elements defining a factor $M$ acting on a Hilbert space $H$ are bounded operators in $B(H)$. Finite type $II_1$ factors are at the root of Murray von Neumann extension of $M$ to include affiliated unbounded operators i.e. those operators that are invariant under any unitary transformation in $M^{'}$. In particular in the type $II_1$ case affiliated operators have a unique resolution of unity ( see point 6 of the introduction of \cite{MvN1} for a clarifying discussion). Affiliated operators will be crucial to define associated reduced density matrices ( see 3.1 of this note).}
\subsection{Dimension and entaglement}
To see the physics meaning of $d_M$ for finite factors more clearly let us start recalling some basic facts about {\it entanglement entropy} for bipartite quantum systems with Hilbert space $H=H_A\otimes H_B$  where both $H_A$ and $H_B$ have finite dimensions $m$ and $n$ respectively. Associated with this system we have a couple of finite type $I$ factors, namely $I_m$ and $I_n$. These factors are just the algebras of bounded operators acting respectively on $H_A$ and $H_B$. Let us denote the two factors $M_A=I_m$ and $M_B=I_n$. Clearly in this case $M_B=M_A^{'}$ and therefore the couple $(M_A,M_B)$ defines a solution to (\ref{problem}).

As it is familiar in quantum mechanics we can associate with any vector state $|\psi\rangle$ in $H$ ( i.e with any pure state of the bipartite system defined by $H=H_A\otimes H_B$ ) the reduced density matrices
\begin{equation}
\rho_A=Tr_B|\psi\rangle\langle \psi|
\end{equation}
and
\begin{equation}
\rho_B=Tr_A|\psi\rangle\langle \psi|
\end{equation}
The Schmidt decomposition allows us to represent any vector state $|\psi\rangle$ as
\begin{equation}
|\psi\rangle =\sum_{i=1,\min(m,n)} \alpha_i (u_i\otimes v_i)
\end{equation}
with $u_i$ and $v_i$ orthonormal sets of vectors in $H_A$ and $H_B$ respectively.
From that  it follows the well known result about von Neumann entropies, namely
\begin{equation}
S_{vN}(\rho_A)=S_{vN}(\rho_B)= \sum_{i=1,\min(m,n)} \alpha_i^2 \log \alpha_i
\end{equation}
for $S_{vN}(\rho)=-Tr(\rho \log \rho)$.
Moreover we see that the {\it maximal entropy} defined by a reduced density matrix associated to a pure vector state $|\psi\rangle$ is,  for both systems, equal to $\log(\min(m,n))$.

Let us now repeat this well known story but using the Murray-von Neumann dimensions $d_{M_A}$ and $d_{M_B}$ for the couple of type $I$ factors associated to the bipartite system. We are interested in characterizing the {\it maximally entangled} states using these dimensions. Following \cite{MvN1} we define
\begin{equation}
d_{M_A}(M_B|\psi\rangle)
\end{equation} 
and
\begin{equation}
d_{M_B}(M_A|\psi\rangle)
\end{equation} 
for $M_{A,B}|\psi\rangle$ the vector spaces generated by $A|\psi\rangle$ for $A\in M_{A,B}$. Note that this {\it relative dimension}  measures, in terms of $d_M$, the {\it $M$-size} of spaces $M^{'}|\psi\rangle$.

Note that for $|\psi\rangle$ unentangled i.e. $|\psi\rangle=|\psi_A\rangle\otimes |\psi_B\rangle$ we get $d_{M_A}(M_B|\psi\rangle)=d_{M_B}(M_A|\psi\rangle)=1$. Moreover for the couple of finite type $I$ factors $(M_A, M_B)$ associated to the bipartite system we get that
\begin{equation}
\max_{|\psi\rangle}d_{M_B}(M_A|\psi\rangle)= \max_{|\psi\rangle}d_{M_A}(M_B|\psi\rangle) = \min(m,n)
\end{equation}
From here it follows that
\begin{equation}\label{relation}
\max_{|\psi\rangle} S_{vN}(\rho_A)= \log (\max_{|\psi\rangle}d_{M_B}(M_A|\psi\rangle))
\end{equation}
and the same for $\rho_B$ just exchanging $A$ by $B$ in the former expression.

Once we have represented the maximal entanglement entropy using the dimension $d_M$ we can
use the properties of $d_M$ to derive some basic properties of the entanglement entropy and to extend some aspects of entanglement entropy to those systems characterized by a couple $(M,M^{'})$ of {\it finite but not normal} type $II_1$ factors.

\subsection{Some basic properties of $d_M$ for finite factors}
For a generic solution $(M,M^{'})$ to (\ref{problem}) we can define $\Delta_0$ and $\Delta^{'}_0$ as the sets of values of $d_M(M^{'}|\psi\rangle$ and of $d_{M^{'}}(M|\psi\rangle)$ when we consider all vector states $|\psi\rangle$ in $H$. For $(M,M^{'})$ a couple $(I_m,I_n)$ of type $I$ factors we easily see that both $\Delta_0$ and $\Delta^{'}_0$ are equal to $\{0,...\min(m,n)\}$. Thus the maximal value of $d_M(M^{'}|\psi\rangle$ and $d_{M^{'}}(M|\psi\rangle$ is $\min(m,n)$ in agreement with the well known result about the maximal entanglement entropy for bipartite finite systems. 

\subsubsection{Murray-von Neumann coupling}
An interesting question addressed in \cite{MvN1} is if we can define a map 
\begin{equation}
\phi:\Delta_0\rightarrow \Delta^{'}_0
\end{equation} 
relating the spectrum $\Delta_0$ and $\Delta^{'}_0$.
The answer of this question is that, for a couple of factors $(M,M^{'})$, such a map exists and it is simply defined by
\begin{equation}
\phi (\alpha) = C(M,M^{'}) \alpha
\end{equation}
for $C(M,M^{'})$ a positive constant {\it independent} of the state $|\psi\rangle$ and $\alpha\in \Delta_0$ \footnote{See \cite{MvN1} theorem 10.} . This implies that

\begin{equation}\label{coupling}
\frac{d_{M}(M^{'}|\psi\rangle)}{d_{M^{'}}(M|\psi\rangle)}=\frac{1}{C(M,M^{'})}
\end{equation}
meaning that the ratio of the relative dimensions is independent of the state $|\psi\rangle$.
For the finite type $I$ case associated with a bipartite system with finite dimensional Hilbert space i.e. for the couple $(M_A,M_B)$ defined above, you can prove that $C(M_A,M_B)=1$. Using (\ref{relation}) this yields to 
$\max S_{vN}(\rho_A)= \max S_{vN}(\rho_B)$ consistently with the fact that $S^{vN}(\rho_A)=S^{vN}(\rho_B)$ for any $\rho$ defined by a pure state $|\psi\rangle$.

For the type $II_1$ case with both $M$ and $M^{'}$ type $II_1$ factors, $C$ has a well defined value that {\it agrees with the inverse of  the  coupling $dim_M(H)$}. For type $III$  as well as for the type $II_{\infty}$ we can fix $C=1$\footnote{Recall that for type $II_1$ the sets $\Delta_0$ and $\Delta_0^{'}$ are both equal to $[0,1]$. For type $III$ both sets are equal to $\{0,\infty\}$. For those cases where $M$ is a type $II_1$ and $M^{'}$ is type $II_{\infty}$ we can fix $C=1$ as well (see \cite{MvN1} theorem 10).}

Let us now briefly recall the definition of $dim_M(H)$ for a type $II_1$ factor $M$ acting on $H$ \footnote{See \cite{Jones} chapter 10 and \cite{Popa} chapter 8.}. Since $M$ is a type $II_1$ factor we can use the trace $\tau_M$ to define the corresponding GNS representation $\pi_{\tau}$ on $H_{\tau}$ with $H_{\tau}$ the GNS Hilbert space defined using the trace $\tau_M$ to define the scalar product. The most natural {\it spatial} property \footnote{The difference between {\it spatial} and {\it algebraic} properties was introduced in \cite{MvN4}. In a nutshell for a von Neumann algebra $M$ algebraic properties are those defining the algebra itself while {\it spatial} properties are those depending on the algebra and the Hilbert space on which the algebra is acting. Thus two algebras $M_1$ and $M_2$ acting on Hilbert spaces $H_1$ and $H_2$ are spatially isomorphic if there exist an isomorphism between $H_1$ and $H_2$ that maps the algebra $M_1$ into $M_2$. As an example where this distinction is relevant think in two solutions $(M_1,M_1^{'})$ and $(M_2,M_2^{'})$ to (\ref{problem}). We can have that although $M_1$ and $M_2$ are algebraically isomorphic and the same for $M_1^{'}$ and $M_2^{'}$ the couples $(M_1,M_1^{'})$ and $(M_2,M_2^{'})$ are not spatially isomorphic. This phenomena will be important for us  in the discussion of black hole evaporation.} we can define is some measure of the {\it relative size} of the Hilbert space $H$  on which $M$ is acting and the GNS Hilbert space $H_{\tau}$. To identify this relative size we need to identify the isometry $u: H\rightarrow H_{\tau}\otimes l^2(\mathbb{N})$ such that $(a\otimes 1) u = u a$ for $a\in M$ and to evaluate $Tr(u^*u)$. This quantity is independent on $u$ and defines the Murray von Neumann coupling as
\begin{equation}
Tr(u^*u)=dim_M(H)
\end{equation}
Using this definition you can easily check that
\begin{equation}
dim_{M}(H_{\tau_M})=1
\end{equation}
and the relation
\begin{equation}
dim_M(H) dim_{M'}(H) =1
\end{equation}
Note that the map $dim_{M}(H)\rightarrow \frac{1}{dim_{M}(H)}$ simply exchange the roles of $M$ and $M^{'}$. In addition for any finite projection $P$ in $M$ we have
\begin{equation}\label{projection}
dim_{PMP}(PH)= dim_{M}(H)\frac{1}{\tau_M(P)}
\end{equation}
For a couple $(M,M^{'})$ of type $II_1$ factors acting on $H$ we can relate the coupling $dim_M(H)$ and the relative dimensions introduced above as
\begin{center}
\boxed{\frac{d_M(M^{'}|\psi\rangle)}{d_{M^{'}}(M|\psi\rangle)}= \frac{1}{C(M,M^{'})}= dim_M(H)}
\end{center}
Using (\ref{projection}) we can define, for a one parameter family $P(\lambda)$ of projections in $M$, the function $F_M(\lambda)$ by
\begin{equation}
F_M(\lambda)=dim_{P(\lambda)MP(\lambda)}(P(\lambda)H)
\end{equation}
This function will be important in our future discussion of the Page curve for finite type $II_1$ factors.
\subsubsection{Entropy deficit and MvN $\theta$ parameter}
For a couple $(M,M^{'})$ of finite factors we can define the deficits $\delta(M)=\frac{d_M(M^{'}|\psi\rangle)}{d_M(H)}$ and $\delta(M^{'})= \frac{d_{M^{'}}(M|\psi\rangle)}{d_{M^{'}}(H)}$. Following \cite{MvN4} ( definition 3.3.1 ) we define
\begin{equation}\label{deficit}
\theta(M,M^{'})= \frac{d_{M^{'}}(H)}{C(M,M^{'})d_{M}(H)}
\end{equation}
which in terms of entropy deficits become $\theta(M,M^{'}) = \frac{\delta(M)}{\delta(M^{'})}$. Note that for a bipartite type $I$ couple $(M_A,M_B)$ we get $\theta(M_A,M_B)=\frac{m}{n}$.

For future use note that although for $(M,M^{'})$ type $I$, type $III$ we have that $d_{M}(M^{'}|\psi\rangle) = d_{M^{'}}(M|\psi\rangle)$ for type $II_1$ we have that 
\begin{equation}\label{couplingrelation}
d_{M}(M^{'}|\psi\rangle) = dim_{M}(H) d_{M^{'}}(M|\psi\rangle)
\end{equation}
In particular this means, in the terminology of \cite{MvN4}, that couples of type $II_1$ factors with different value of the coupling are not {\it spatially} ( see footnote 7) isomorphic \footnote{This relation will be crucial in our algebraic discussion of black hole evaporation}. 
\section{The Page curve and the central dogma.}

\subsection{Setup and notation: the QES prescription}
In order to describe the process of black hole evaporation we will define, at any time $t$, two systems to be denoted $B_t$ and $R_t$. $B_t$ should account for the {\it degrees of freedom} needed to describe the black hole from the outside
while $R_t$ will represent the degrees of freedom needed to describe the radiation evaporating into a reservoir. Let us denote $W(B_t)$ and $W(R_t)$ the associated {\it entanglement wedges}. Before going ahead let us briefly review the definition and meaning of entanglement wedges.

In the holographic setup the entanglement wedge of a boundary region $A$ is defined as the bulk region $H(A)$ that is holographically dual to $A$. This means that local bulk operators in the entanglement wedge $H(A)$ can be {\it reconstructed} in terms of boundary local operators with support in $A$. The RT,HRT,EW \cite{RT,HRT,EW} prescription to identify $H(A)$ is defined as follows. First of all you look for bulk surfaces $\sigma(A)$ homologous to $A$ and you define  $H(A,\sigma(A))$ as the homology hypersurface satisfying $\partial H(A,\sigma)=\sigma(A)\cup A$. Next you define a generalized entropy functional \cite{EW} 
\begin{equation}
S_{gen}(H(A),\sigma(A))=\frac{Area(\sigma(A))}{4G_N} + S_{vN}(\rho_{H(A,\sigma(A))})
\end{equation}
 for $\rho_H$ {\it some reduced density matrix} on $H(A,\sigma(A))$ \footnote{Generically this von Neumann entropy is interpreted as the von Neumann entropy of the quantum field fluctuations in $H(A)$. At this qualitative level we refer to this entropy as the von Neumann entropy of {\it some} reduced density matrix to be defined, once we count with some microscopic model.}. Finally you extremize this generalized entropy functional by embedding $\sigma(A)$ into the bulk geometry and requiring the {\it quantum mean curvature} \cite{Wall}, defined by the generalized entropy, to vanish. The solution $\sigma(A)$ to this process is known as the QES and the corresponding value of $S_{gen}(H(A),\sigma(A))$ is identified with the entanglement entropy $S_{vN}(\rho_A)$ of the boundary region $A$. This entropy is the von Neumann entropy for a reduced density matrix $\rho_A$ on the boundary region $A$. It is important to remark that in this beautiful construction there are several pieces that require a more precise definition. In particular how we define the reduced density matrices $\rho_H$ and $\rho_A$ as well as how we define the traces involved in the definition of the corresponding von Neumann entropies. We will extensively discuss those issues in this note but before let us export the former construction to the case of black hole evaporation ( see for a nice review and references \cite{Malda}).

 For the case of the black hole evaporation problem and if we don't assume any holographic description, we will define the entanglement wedges $W(B_t)$ and $W(R_t)$, at a given time $t$, by analogy with the definition used in the holographic setup.

In order to do that we  define a foliation, of the black hole space time geometry, in terms of Cauchy hipersurfaces, that we will denote $\Sigma_t$. At any time $t$ we can define a codimension two Cauchy {\it splitting surface} $\chi(t)$ \cite{EW4} ( the analog of $\sigma$ ) such that $\Sigma_t =\Sigma_B(t)\cup \chi(t) \cup \Sigma_R(t)$ and we will define, relative to this splitting surface, the entanglement wedges $W(B_t)$ and $W(R_t)$ as the causal developments of $\Sigma_B(t)$ and $\Sigma_R(t)$ ( see Figure 1 and Figure 2). 

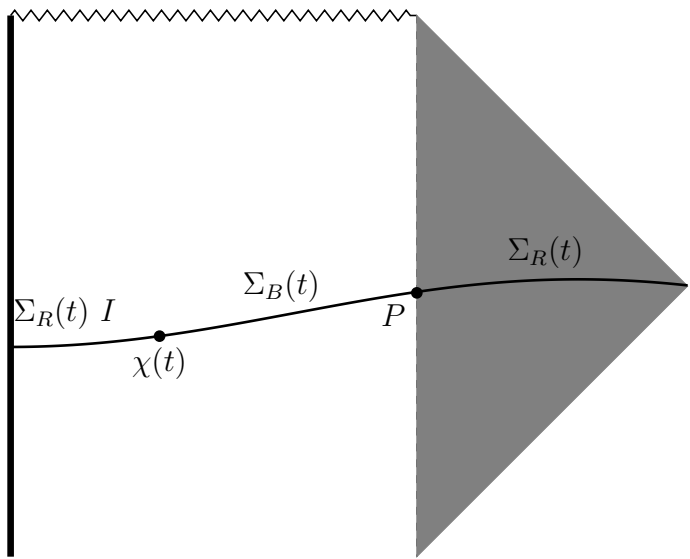
\begin{figure}
    \begin{center}
        \begin{tikzpicture}[scale=1.8]
            \draw[line width=2.5](-1,2)--(-1,-2);
            \draw[dashed](2,2)--(2,-2);
            \draw[singularity](-1,2)--(2,2);
            \filldraw[gray,opacity=0.1] (2,2)--(4,0)--(2,-2)--cycle;
            \draw[domain=0:1.592*pi, samples=100, shift={(-1,-.2)}, line width = 1] plot (\x, {.25*cos(deg(.75*\x-3*pi))});
            \draw[thick]  (-0.6,-.4)circle(0pt) node[left,above] {$\Sigma_R(t)$ $I$};
            \filldraw[thick]  (0.1,-.37)circle(1pt) node[below] {$\chi(t)$};
            \draw[thick]  (1,-0.2)circle(0pt) node[left,above] {$\Sigma_B(t)$};
            \filldraw[thick]  (2,-0.05)circle(1pt) node[anchor=north east] {$P$};
            \draw[thick]  (2.95,0.05)circle(0pt) node[above] {$\Sigma_R(t)$};

        \end{tikzpicture}
    \end{center}
    \caption{Representation of the QES $\chi(t)$ and the corresponding island $I$.}\label{fig:1} 
\end{figure}
 
 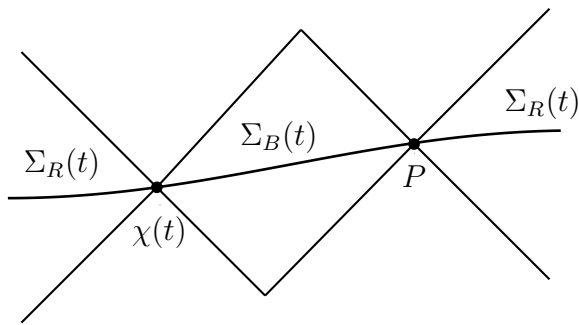
\begin{figure}
    \begin{center}
        \begin{tikzpicture}[scale=1.8]
            \draw[domain=0:1.3*pi, samples=100, shift={(-1,-.2)}, line width = 1] plot (\x, {.25*cos(deg(.75*\x-3*pi))});
            \draw[thick]  (-0.6,-.4)circle(0pt) node[left,above] {$\Sigma_R(t)$};
            \filldraw[thick]  (0.1,-.37)circle(1pt) node[below] {};
            \filldraw[thick]  (0.12,-.5)circle(0pt) node[below] {$\chi(t)$};
            \draw[thick]  (1,-0.2)circle(0pt) node[left,above] {$\Sigma_B(t)$};
            \filldraw[thick]  (2,-0.05)circle(1pt) node[anchor=north] {};
            \filldraw[thick]  (2,-0.12)circle(0pt) node[below] {$P$};
            \draw[thick]  (2.95,0.05)circle(0pt) node[above] {$\Sigma_R(t)$};
            \draw[thick](0.1,-.37)--(-.9,1-.37);
            \draw[thick](0.1,-.37)--(-.9,-1.37);
            \draw[thick](0.1,-.37)--(1.165,1.165-.37);
            \draw[thick](0.1,-.37)--(.9,-0.8-.37);
            \draw[thick](1.165,1.165-.37)--(2,-0.05);
            \draw[thick](.9,-0.8-.37)--(2,-0.05);
            \draw[thick](2,-0.05)--(2+1,-0.05+1);
            \draw[thick](2,-0.05)--(2+1,-0.05-1);
        \end{tikzpicture}
    \end{center}
    \caption{Entanglement Wedges $W_B(t)$ and $W_R(t)$.}\label{fig:2} 
\end{figure}

At any time $t$ and for any Cauchy splitting surface $\chi(t)$ we define the {\it generalized entropy functional} as
\begin{equation}\label{gen}
S_{gen}(\chi(t)) = \frac{Area(\chi(t))}{4G_N} + S_{vN}(\rho_{\Sigma_B(t)})
\end{equation}
where $\rho_{\Sigma_B(t)}$ is assumed to be a reduced density matrix on $W(B_t)$. We can equally define (\ref{gen}) using a reduced density matrix $\rho_{\Sigma_{R(t)}}$ on $W(R_t)$. We need to assume that both generalized entropies are identical. 

Next you extremize $S_{gen}$ on the space of splitting surfaces $\chi(t)$ \footnote{Using the analog of the quantum mean curvature condition for (\ref{gen}).} and identify the solution as the black hole QES, let us say $\chi_{QES}(t)$. Finally you identify the so defined extremal value of $S_{gen}(\chi_{QES}(t))$ as the von Neumann entropy of the black hole at time $t$. 

This QES representation of the black hole entropy can explain several physics aspect of the time dependence of the black hole entropy. In particular we expect $S_{gen}(\chi_{QES}(t))$ to follow a Page curve. Moreover we could interpret the appearance, after Page time, of solutions $\chi_{QES}(t)$ with non vanishing $Area(\chi_{QES}(t))$ as implying the expected reconstruction, in agreement with Hayden-Preskill conjecture \cite{HP}, of parts of the black hole interior in terms of asymptotic radiation modes. 

However, as said, this construction requires a more precise understanding of the reduced density matrices as well as the traces involved in the definition of $S_{gen}(\chi(t))$.

\subsection{The algebraic picture and the central dogma}
As discussed the QES prescription allows us to define at each time $t$ the entanglement wedges $W_B(t,\chi^{QES}(t))$ and $W_R(t,\chi^{QES}(t))$ where we make explicit the dependence of these wedges on the QES $\chi^{QES}(t)$.

Next let us figure out the algebraic description of this construction. The algebraic approach assumes that associated to the two entanglement wedges you can define two von Neumann algebras $M_B(t)$ and $M_R(t)$ acting on a Hilbert space $H$ such that at any $t$
\begin{equation}\label{commutant}
M_B(t)=(M_R(t))^{'}
\end{equation}
This is equivalent to say that $B(H)=M_B(t)\otimes M_R(t)$ and that $(M_B(t),M_R(t))$ is a couple of factors satisfying (\ref{problem}) i.e. $M_B(t)\cap M_R(t) ={\mathbb{C}}1$.

The strong version of {\it The central dogma} (see \cite{Malda}) imposes a priori constraints on the nature of these algebras. Indeed, the dogma requires to assume that $H$ is {\it finite dimensional}. This automatically implies that $M_B(t)$ and $M_R(t)$ are finite type $I$ factors, let us say $I_{m_B(t)}$ and $I_{m_{R}(t)}$, with $m_{B}(t),m_{R}(t)$ finite integer numbers satisfying $m_{B}(t)m_{R}(t)=N$ for $N$ the {\it finite} dimension of $H$. Moreover this implies the existence of a bipartite decomposition $H=H_B(t)\otimes H_R(t)$ with $M_B(t)=B(H_B(t))$ and $M_{R}(t)=B(H_R(t))$ \footnote{Note that this assumption on the finiteness of the Hilbert space prevents us to model the algebras $M_B(t)$ and $M_R(t)$ as algebras of local operators with support on the corresponding entanglement wedges. Such model of $M_B(t)$ and $M_R(t)$ will imply that these algebras are type $III_1$ \cite{Araki} ( see also section 12 of \cite{AW}). A priori this is not necessarily a problem if we assume that the type $I$ factors $M_B(t)$ and $M_R(t)$ describe some abstract observables built in terms of some abstract q-bits able to encode the full information ( from the outside ) about the black hole.}.

In these conditions we can define for any {\it vector state} $|\psi\rangle \in H$ the von Neumann entropy $S^{vN}(|\psi\rangle,t)= Tr_{H_B(t)}(\rho_R(t)) = Tr_{H_R(t)}(\rho_{B}(t))$ with $\rho_{R,B}(t)= Tr_{H_B(t),H_R(t)}|\psi\rangle\langle\psi|)$ as well as the maximal entropy 
 
\begin{equation}
S^{vN}(t)= \max_{|\psi\rangle\in H}S^{vN}(|\psi\rangle,t)
\end{equation}
that we expect will correspond to the physical entropy of the black hole.
Obviously in this type $I$ version of the strong central dogma we get for this maximal von Neumann entropy
\begin{equation}
S^{vN}(t)=\log (\min(m_B(t),m_R(t)))
\end{equation}
The maximal fine grained entropy $S^{vN}(t)$ defined above has, as discussed in the previous section, a nice algebraic meaning as a relative dimension, namely
\begin{equation}
S^{vN}(t)= \log(\max_{|\psi\rangle}(d_{M_B(t)}(M_R(t)|\psi\rangle)))
\end{equation}
i.e. is the $\log$ of the maximal value of $d_{M_B(t)}(M_R(t)|\psi\rangle)$ on the space of vector states in $H$. Moreover in this finite type $I$ version we have $d_{M_B(t)}(M_R(t)|\psi\rangle) = d_{M_R(t)}(M_B(t)|\psi\rangle)$.

\subsubsection{Coarse grained input and type $I$ Page curve}
In order to constraint the time dependence of $S^{vN}(t)$ we can use as input the time dependence of two coarse grained entropies, namely $S^{th}_{bh}(t)$ defining the thermodynamic Bekenstein Hawking entropy of the black hole at each instant of time and $S^{Hawking}_{rad}(t)$ defined by the semiclassical computation of Hawking of the entanglement entropy of the radiation. The dependence on time of these coarse grained entropies is represented in Figure 3. Let us denote $t_P$ the value of time at which these coarse grained entropies are identical. 

Since these entropies are coarse grained we can impose that 
\begin{equation}\label{ineq1}
S^{vN}(t) \leq S^{Hawking}_{rad}(t)
\end{equation}
 as well as 
 \begin{equation}\label{ineq2}
 S^{vN}(t) \leq S^{th}_{bh}(t)
 \end{equation}
for any value of $t$. This implies that $S^{vN}(t)$ is represented by some curve $P$ in Figure 3. We will refer to these types of curves for $S^{vN}(t)$ as {\it Page curves}.
 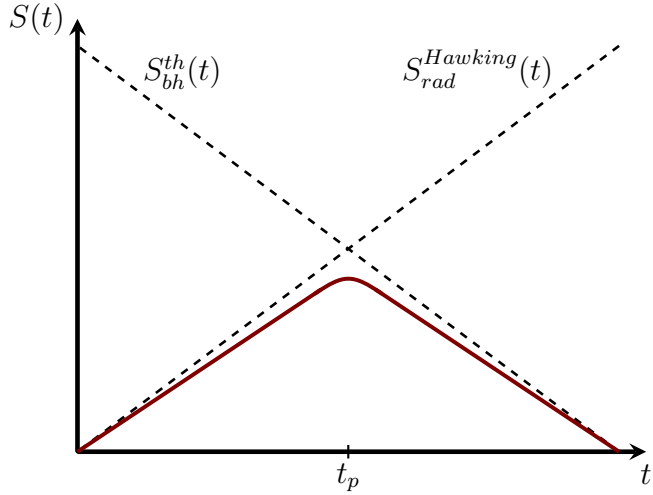
\begin{figure}
    \begin{center}
        \begin{tikzpicture}[scale=1.80]
        \draw[-stealth,thick,line width=1.75](0,0)--(4.2,0)node[anchor=north]{$t$};
        \draw[-stealth,thick,line width=1.75](0,0)--(0,3.2)node[anchor=east]{$S(t)$};
        \draw[thick,line width=1,dashed](4,3)--(0,0);
        \draw[thick,line width=1,dashed](4,0)--(0,3);
        \filldraw(0,0) circle(.25pt);
        \draw[thick]  (4-.4,3-.4)circle(0pt) node[anchor=south east ] {$S^{Hawking}_{rad}(t)$};
        \draw[thick]  (.8-.4,3-.4)circle(0pt) node[anchor=south west ] {$S^{th}_{bh}(t)$};
        \draw[thick]   (2,-0.05) --(2,0.05);
        \draw[thick]  (2,0)circle(0pt) node[anchor=north ] {$t_p$};
        \draw[mydarkred,line width=1.5]  (0,0)--(1.8,1.2) node[anchor=north ] {};
        \draw[mydarkred,line width=1.5]  (4,0)--(2.2,1.2) node[anchor=north ] {};
        \draw[mydarkred,line width=1.5] (1.75,1.166)..controls+(.22,.15) and +(-.22,.15)..(2.25,1.167);
        \end{tikzpicture}
    \end{center}
    \caption{Curves representing the coarse grained input $S^{th}_{bh}(t)$ and $S^{Hawking}_{rad}(t)$. In color a Page curve for $S^{vN}(t)$.}\label{fig:3} 
\end{figure}

The simplest type $I$ representation of the so defined coarse grained entropies could be
\begin{equation}
S^{th}_{bh}(t)= \log (d_{M_{B}(t)}(H))
\end{equation}
and
\begin{equation}
S^{Hawking}_{rad}(t)=\log (d_{M_{R}(t)}(H))
\end{equation}
We can now evaluate the  deficits defined as $\delta_{bh}(t)=S^{th}_{bh}(t)-S^{vN}(t)$ and $\delta_{rad}(t)= S^{Hawking}_{rad}(t)-S^{vN}(t)$. For type $I$ we immediately get
\begin{equation}
\delta_{bh}(t)=\delta_{rad}(t) +\log \theta_{MvN}(t)
\end{equation}
for $\theta_{MvN}(t)$ defined in (\ref{deficit}). Note that in this type $I$ picture the time $t_P$ is fixed by the condition $\theta(t_P)=1$. 

The simplest {\it type $I$} Page curve saturating the inequalities (\ref{ineq1}) and (\ref{ineq2}) is obviously given by $S^{vN}(t)=\log \min(m(t),n(t))$ with $n(t)=\min(m(t),n(t))$ for $t<t_P$ and with $m(t)= \min(m(t),n(t))$ for $t>t_P$. In summary once we accept the strong version of the central dogma and consequently we associate with the entanglement wedges of the black hole and the radiation, finite type $I$ factors, the Page curve follows from the coarse grained input. Is this the end of the story ? Very likely the answer is negative since in this type $I$ version we miss all the information about generalized entropy as well as the underlying mechanism of purification

Hence the natural question should be: What is the relation between the type $I$ von Neumann entropy $S^{vN}(t,|\psi\rangle)$ defined above and the QES generalized entropy $S_{gen}(t,\chi^{QES}(t))$ ? Note that contrary to the simple algebraic type $I$ description the semiclassical QES prescription explains the underlying dynamics leading to the Page curve.

\subsection{Type $I$ and QES}
To establish the relation between the type $I$ definition of von Neumann entropy and the QES generalized entropy we need to find a relation of the type
\begin{equation}\label{basicrelation1}
S^{vN}(\rho_{R(t,\chi^{QES}(t))}(|\psi\rangle)= \frac{Area(\chi^{QES}(t))}{4G_N} + S^{vN}(\hat \rho_{R(t,\chi^{QES}(t))})
\end{equation}
where $\rho_{R(t,\chi^{QES}(t))}(|\psi\rangle)$ is defined as $tr_{H_R(t)}|\psi\rangle\langle\psi|$ and where $\hat \rho_{R(t,\chi^{QES}(t))}$ is the reduced density matrix used in the definition of the generalized QES entropy.

Clearly $\hat \rho_{R(t,\chi^{QES}(t))}$, in this type $I$ version where the Hilbert space factorizes, must be the reduced density matrix defined by some state $|\psi\rangle$ most likely the vacuum. Irrespectively what state we use this reduced density matrix should be, in the context of type $I$, where we factorize the Hilbert space as $H=H_B(t)\otimes H_R(t)$, equal to the reduced density matrix $\rho_{R(t,\chi^{QES}(t))}(|\psi\rangle)$ appearing in the l.h.s of (\ref{basicrelation1}) provided we use the {\it same state} $|\psi\rangle$. But this implies that the piece $\frac{Area(\chi^{QES}(t))}{4G_N}$ in (\ref{basicrelation1}) must be zero.

In other words we conclude that:

\vspace{0.3 cm}

{\it For $(M_B(t),M_R(t))$ a couple of finite type $I$ factors, which is compulsory if $H$ is finite dimensional, there are not QES solutions with $Area(\chi(t))$ non vanishing.}

\vspace{0.3 cm}

Therefore for finite dimensional Hilbert space QES islands with non vanishing value of $\frac{Area(\chi^{QES}(t))}{4G_N}$ don't exist.

\vspace{0.3 cm}

To avoid this conclusion we need to relax the condition that $(M_B(t),M_R(t))$ is a couple of type $I$ factors. The simplest way to relax this condition is to assume that {\it the Hilbert space $H$ is infinite dimensional}.  

\section{A {\it finite factor version} of the central dogma}
The simplest and more natural generalization of the central dogma could be to impose that:

\vspace{0.3 cm}

{\it We have at each time $t$ a couple of {\bf finite} factors $(M_B(t),M_R(t))$ satisfying (\ref{commutant}) but acting on a separable Hilbert space $H$ of {\bf infinite dimension}.}

\vspace{0.3 cm}

In other words we keep finiteness of the factors associated to the two entanglement wedges {\it but} acting on an infinite dimensional Hilbert space. In this case the factors $M_B(t)$ and $M_R(t)$ cannot be normal i.e. $H$ does not factorize.

The only possibility in this case is that these factors are finite type $II_1$ factors. 

\subsection{von Neumann entropies and affiliated operators}
Before attempting to define a Page curve for black hole evaporation using finite type $II_1$ factors it will be convenient to discuss how for these finite factors we can define the analog of von Neumann entropy.

Factors that are not type $I$ are not normal, therefore for a couple $(M,M^{'})$ acting on a Hilbert space $H$ we cannot define the corresponding factorization of the Hilbert space. At first sight this factorization problem seems to imply that we cannot give meaning to the reduced density matrices we normally use to define von Neumann entanglement entropies. 
A way to avoid this problem for factors having a trace state is to use as reduced density matrices self adjoint {\it affiliated operators}. Next we will briefly review this construction in general terms ( see also footnote 3 ).

Let us start with a von Neumann algebra $M$ with a finite or semifinite faithful normal trace. Let us generically denote $\tau_M$ this trace state i.e. $\tau_M$ is a linear form on $M$ satisfying the trace property $\tau_M(ab)=\tau_M(ba)$ for $a,b$ arbitrary elements of $M$. Let us now consider a generic state $\Phi$ on $M$ i.e. a linear form on $M$ that is not satisfying the trace property. To be more precise you can consider a weight on $M$ \footnote{By weight on $M$ we will mean faithful, normal and semifinite. Recall that a weight $\Phi$ on $M$ is defined as a linear form on $M^{+}$ i.e. $\Phi(a+b)=\Phi(a)+\Phi(b)$ and $\Phi(\lambda a)=\lambda\Phi(a)$ for $a\in M^{+}$ such that for any $a\in M^{+}$ we have that $\Phi(a)=0$ implies $a=0$ and such that the set of $a\in M^{+}$ for which $\Phi(a^{*}a)$ is finite is dense in $M$ ( see \cite{Connes} for definitions).}. Now we can define the affiliated self adjoint density matrix $\rho_{\Phi}$ by
\begin{equation}\label{density}
\Phi(a)=\tau_M(\rho_{\Phi}a)
\end{equation}
In standard quantum mechanics and for $\Phi$ a vector state i.e. a pure state $|\Phi\rangle$ , the definition (\ref{density}) implies the familiar relation $\rho_{\Phi}=|\Phi\rangle\langle\Phi|$. In particular for $M$ the type $I$ factor defined as $B(H)$ for $H$ a finite dimensional Hilbert space and for $\tau_M$ the standard trace this is the usual quantum mechanical definition of the density matrix associated to a pure state. 

Note that in abstract terms the definition of the affiliated self adjoint operator $\rho_{\Phi}$ only depends on the existence of a semifinite trace state on $M$. Now we are interested in an explicit expression of $\rho_{\Phi}$. Recall that for $\Phi$, that we assume is a faithful, normal and semifinite weight, we can define the GNS isomorphism $\pi_{\Phi}$ of $M$ into bounded operators of the GNS Hilbert space $H_{\Phi}$. Using Tomita-Takesaki theoy we can define the involution $J_{\Phi}$ as well as the modular operator $\Delta_{\Phi}$. The associated modular automorphism $\sigma_t^{\Phi}$ is defined by $\pi_{\Phi}(\sigma_t^{\Phi} (a)) = \Delta_{\Phi}^{it} \pi_{\Phi}(a) \Delta_{\Phi}^{-it}$. Finally for two states $\Phi$ and $\Psi$ we can define Connes cocycle \cite{Connes} $u_{\Phi,\Psi}(t)$ by
\begin{equation}
\sigma_t^{\Psi} = u_{\Phi,\Psi}(t) \sigma_t^{\Phi} u_{\Phi,\Psi}^{*}(t)
\end{equation}
Now we can use these ingredients to identify $\rho_{\Phi}$. Note that the trace $\tau_M$ defines a weight on $M$ so we can {\it compare} the weights $\Phi$ and the one defined by $\tau_M$ and evaluate the corresponding Connes cocycle \cite{Connes}, let us say $u_{\Phi,\tau_M}(t)$. The
 key result defining the affiliated $\rho_{\Phi}$ is ( see \cite{CT}):
 \begin{equation}\label{formula}
 \rho_{\Phi}^{it}= u_{\Phi,\tau_M}(t)
 \end{equation}
 Once we have identified $\rho_{\Phi}$ we can define the von Neumann entropy \cite{Longo}
 \begin{equation}
 S^{vN}(\Phi)=- \tau_{M}(\rho_{\Phi}\log\rho_{\Phi})
 \end{equation}
But, What is the physics meaning of this von Neumann entropy for factors that are not normal? To address this question let us start with the case of the eternal AdS black hole.
 
\subsection{The case of the eternal AdS black hole} 
 The von Neumann entropy defined in the previous section using the affiliated density matrix $\rho_{\Phi}$ has been worked out in \cite{Witten2} for the particular case of the two sided AdS black hole. 
 
 The setup in \cite{Witten2} is defined by a couple of type $III_1$ factors $({\cal{A}}_{r},{\cal{A}}_{l})$ with ${\cal{A}}_{r}=({\cal{A}}_{l})^{'}$. If we assume that these factors are hyperfinite, them both are algebraically isomorphic \cite{haagerup} to Araki-Woods factor $R_{\infty}$ \cite{AW}. Thus, to simplify the notation we can refer to both type $III_1$ factors simply as ${\cal{A}}$. Algebraically we can work with the isomorphic factor $\tilde{\cal{A}}={\cal{A}}\otimes F_{\infty}$ with $F_{\infty}$ the $I_{\infty}$ factor of bounded operators acting on the Hilbert space $L^2(\mathbb{R})$. For the time being we don't need to be more explicit about the physics meaning of this $F_{\infty}$ factor. The reader familiar with \cite{Witten1} and \cite{Witten2} will identify this $F_{\infty}$ factor with an external {\it observer} characterized by one quantum degree of freedom and Hilbert space of states $L^2(\mathbb{R})$ i.e. as $F_{\infty}=B(L^2(\mathbb{R})$. Let us postpone the discussion on the physics role of this added observer. What is important is that for the type $III_1$ factor ${\cal{A}}$ that we are using we can always use the {\it algebraically isomorphic} representation $\tilde{\cal{A}}$.

The factor $\tilde{\cal{A}}$ has no any semifinite trace state so in order to define an affiliated density matrix we need first to define a trace state \footnote{The mathematical reason to work with $\tilde{\cal{A}}$ isomorphic to ${\cal{A}}$ is to use {\it generalized traces} as defined in 4.3 of \cite{Connes}. Given a factor ${\cal{A}}$ we can define on $\tilde{\cal{A}}$ a generalized trace, let us say $\omega$, such that the couple of centralizers $(\tilde{\cal{A}}_{\omega},\tilde{\cal{A}}_{\omega}^{'})$ defines a solution to (\ref{problem}). See theorem 4.2.6 of \cite{Connes}.}
. This can be done in an almost ( algebraically) unique way using a {\it dominant weight} \cite{CT} $\omega$ on $\tilde{\cal{A}}$. This dominant weight is defined as $\phi\otimes \bar\omega$ for $\phi$ a weight on ${\cal{A}}$ and $\bar\omega$ a weight on $F_{\infty}$ defined as 
\begin{equation}
\bar\omega(x)= Tr(\tilde \rho x)
\end{equation}
with $Tr$ the standard trace on the $F_{\infty}$ factor $B(L^2(\mathbb{R}))$ and with $\tilde \rho$ defined by the generator of translations in $L^2(\mathbb{R})$ i.e. $\tilde \rho^{it}= U_t$ with $U_t f(x)=f(x+t)$ for $f(x)\in L^2(\mathbb{R})$. Note that in the notation of \cite{Witten2} the generator of $U_t$ is identified with {\it the observer hermitian, but not positive, formal hamiltonian}. The dominant weight $\omega$ satisfies the KMS condition and consequently we can define a semifinite trace $\tau_{\omega}$ on the centralizer $\tilde{\cal{A}}_{\omega}$ \footnote{Recall that the centralizer is defined by those elements in $\tilde{\cal{A}}$ invariant under the action of the corresponding modular automorphism.}. This centralizer is a type $II_{\infty}$ factor isomorphic to the crossed product 
\begin{equation}\label{crossedw}
{\cal{A}}\rtimes_{\sigma_{\phi}} \mathbb{R} 
\end{equation}
Now, and relative to a fixed dominant weight $\omega$, we can use the former construction to define the affiliated density matrices for any weight $\Psi$ on $\tilde{\cal{A}}_{\omega}$ by
\begin{equation}\label{afiliated}
\Psi(a)=\tau_{\omega}(\rho_{\Psi} a)
\end{equation}
for any $a\in \tilde{\cal{A}}_{\omega}$. In \cite{Witten2} this affiliated density matrix was evaluated, for some special states $\Psi$ ( the so called semiclassical states ), as well as the corresponding von Neumann entropy $S^{vN}(\Psi)$.
 
 A very nice result of \cite{Witten2} was to interpret the so defined von Neumann entropy as a {\it generalized entropy}\footnote{See also \cite{Kudler}.}. In simple cases we can formally figure out the origin of this result.
As a concrete example let us define the dominant weight $\omega=\phi\otimes \bar\omega$ for a concrete weight $\phi$ on ${\cal{A}}$ and let us assume that $\Psi$ is defined as $\Psi=\psi\otimes \bar\omega^{'}$ for $\psi$ a different weight on ${\cal{A}}$ and $\bar\omega^{'}$ also a different weight on the $F_{\infty}$ factor. What is the density matrix $\rho_{\Psi}$ relative to $\tau_{\omega}$? 

In this case this density matrix could be directly derived from the general result (\ref{formula}). Formally it can be represented as
\begin{equation}\label{cocycle}
\rho_{\Psi}^{it}= u_{\psi,\phi}(t) C_{\bar \omega^{'}, \bar\omega}(t)
\end{equation}
where $u_{\psi,\phi}(t)$ defines the cocycle between the two states $\phi$ and $\psi$ of ${\cal{A}}$ and where $C_{\bar \omega^{'},\bar\omega}(t)$ represents the cocycle between the  weights $\bar\omega^{'}$ and $\bar\omega$  of the $F_{\infty}$ factor i.e. {\it the observer factor} \footnote{The relation (\ref{cocycle}) corresponds to expression (3,20) in \cite{Witten2}.} 

Expression (\ref{cocycle}) captures the generalized entropy meaning of the von Neumann entropy defined by 
$\rho_{\Psi}$ \footnote{The {\it area} contribution to the generalized entropy being determined by the cocycle $C_{\bar \omega^{'},\bar{\omega}}$.}.

Now we can come back to the puzzle created by the relation (\ref{basicrelation1}) relating the von Neumann entropy with the generalized entropy function used to define the QES prescription. The answer is that once we work with the type $II_{\infty}$ factor $\tilde{\cal{A}}_{\omega}$ the von Neumann entropy defined by the affiliated density matrices is a generalized entropy \cite{Witten2}.


{\bf Remark } In \cite{Witten2} the dominant weight $\omega$ used to define a trace on the corresponding centralizer ${\tilde{\cal{A}}}_{\omega}$ is defined as $\Psi\otimes \bar\omega$ with $\Psi$ a state on ${\cal{A}}$ formally interpreted as defining the large $N$ limit of a TFD state and with $\bar\omega$ the weight on $F_{\infty}$ defined above. Using the crossed product representation ${\tilde{\cal{A}}}_{\omega}={\cal{A}}\rtimes_{\sigma_{\Psi}}{\mathbb{R}}$ and using the corresponding conditional expectation $E_{\Psi}$ we can generate any element $a\in{\tilde{\cal{A}}}_{\omega}$ in terms of the element $E_{\Psi}a$ and some function $f_a(x)$ on the generator $x$ of $U_t$ i.e.  the {\it observer formal hamiltonian}. Using this representation i.e. $a=E_{\Psi}a\otimes f_a(x)$ we define the trace $Tr(a)$ induced by $\omega$ as $\Psi(E_{\Psi} a)\bar \omega(f_a(x))=:\langle \Psi|E_{\Psi} a|\Psi\rangle \bar\omega(f_a(x))$. Note that for the case where $E_{\Psi} a=\hat a$ (in the notation of \cite{Witten2}) and for $f_a(x)=1$ we get the trace defined in \cite{Witten2} ( see equation 2.24) \footnote{In \cite{Witten2} the weight $\bar\omega$ is represented as $\bar\omega(\cdot)=Tr(e^{\beta x} \cdot)$. We are using the convention where $\beta x$ is identified with the dimensionless  generator of $U_t$ that we simply denote as $x$. Note that the limit $\beta=0$ is not defining a dominant weight ( see theorem 1.3 of \cite{CT}).}. 

\subsubsection{Takesaki duality}
Takesaki duality \cite{Takesaki} leads to an illuminating crossed product representation of the type $III_1$ factor $\tilde{\cal{A}}$ as
\begin{equation}\label{crossed}
\tilde{\cal{A}}=\tilde {\cal{A}}_{\omega}\rtimes_{\theta_s}{\mathbb{R}}
\end{equation}
with $\theta_s$ the automorphism of $\tilde {\cal{A}}_{\omega}$ scaling the semifinite trace $\tau_{\omega}$ as $\theta_s \tau_{\omega}=e^{-s}\tau_{\omega}$.

The representation (\ref{crossed}) reveals the generators of $\tilde{\cal{A}}$ as: i) The algebra of {\it the gravitationally dressed operators} that we will denote $\tilde{\cal{A}}_d$ and that can be defined as $E(\tilde{\cal{A}})$ for $E$ the conditional expectation associated to the crossed product (\ref{crossedw}), ii) The generator $U_t$ used to define the $\bar\omega$ and iii) The generator $V_s$ defining the Weyl representation of the canonical commutation relation $U_tV_s=e^{ist}V_sU_t$ on $L^2(\mathbb{R})$. Note that the generator $V_s$ acts on the dominant weight $\omega$ as $V_s\omega =e^{-s}\omega$ i.e. implements a scale transformation of the trace $\tau_{\omega}$. In summary the representation (\ref{crossed}) shows that the type $III_1$ factor $\tilde{\cal{A}}$ is generated by the gravitationally dressed operators and {\it the observer quantum algebra} defined by the Weyl representation $U_tV_s=e^{ist}V_sU_t$.

In addition it is easy to see that the von Neumann entropy $S^{vN}(\Psi)$ for any weight $\Psi$ on $\tilde{\cal{A}}_{\omega}$ defined using the affiliated density matrices associated to $\tau_{\omega}$ change under the action of the automorphism $\theta_s$ as
\begin{equation}
S^{vN}(\Psi)\rightarrow S^{vN}(\Psi)+s
\end{equation}
\subsection{The importance of {\it spatial} properties}

Summarizing in this section we have discussed solutions to (\ref{problem}) of the type $(\tilde{\cal{A}}_{\omega},\tilde{\cal{A}}_{\omega}^{'})$ for $\omega$ a dominant weight on $\tilde{\cal{A}}$. For any dominant weight $\omega$ the factors $\tilde{\cal{A}}_{\omega}$ and $\tilde{\cal{A}}_{\omega}^{'}$ are algebraically isomorphic to the unique hyperfinite type $II_{\infty}$ factor. 

As already pointed out any description of the process of black hole evaporation requires to use a one parameter family of solutions to (\ref{problem}) in terms of factors $M_B(t)$ and $M_R(t)$ associated to the entanglement wedges of the black hole and the radiation. One of the key ingredients of the proposal of this paper, as we will discuss in the rest of this note, is that the couples of factors $(M_B(t),M_R(t))$ describing the evaporation process {\it cannot be spatially isomorphic at different times of the evaporation process}.

For the eternal AdS black hole we can fix a couple of factors describing the left and right wedges. In terms of these 
factors, uniquely determined once we have fixed a concrete dominant weight, we can define, relative to the corresponding trace, affiliated density matrices for different states. These states can be used to describe different physical configurations on the background of the eternal AdS black hole. By contrast when we are describing the physical process of evaporation we need, as indicated by the QES prescription, to use different couples $(M_B(t),M_R(t))$ for different times. To identify the dynamics of the process requires to identify the {\it differences} between these couple of factors, in particular the differences giving rise to the Page curve. As we will argue in the next sections these differences should be, in Murray von Neumann terminology, {\it spatial}.

\section{Black hole evaporation: preliminary comments}
To start with let us pose the following general question: What should be the minimum conditions that we should demand must satisfy an algebraic description of the evaporation of a black hole?

A reasonable answer, motivated by the QES prescription, should involve the following  requirements: 

i) A one parameter family of couples of factors $(M_B(s),M_R(s)$ with a well defined semifinite trace $\tau(s)$ satisfying (\ref{problem}) for any value of $s$ \footnote{At this point $s$ is just a formal parameter distinguishing different moments of the evaporation process. Its relation to a physical time will be discussed later.} . 

ii) A finite entropy function $S(s)$ defined on the basis of $\tau(s)$ satisfying Page's curve.

iii) To identify a physics interpretation of $S(s)$ to explain how during the evaporation process the information, initially contained inside the black hole, is transferred and encoded into the state describing  the final radiation.

Before going into the description of some possible models let us briefly comment on the meaning of these requirements. If we rule out the already discussed possibility of using finite type $I$ factors we should assume that the couples of factors $(M_B(s),M_R(s))$ are acting on {\it an infinite dimensional Hilbert space}. This implies that these factors are not {\it normal}. This could be interpreted as a necessary requirement to have a semiclassical space-time picture of the process in a connected space-time \footnote{As stressed in \cite{EL} factorizability of the Hilbert space implies some lack of space-time connectivity.}. To require the existence of at least a semifinite trace  is necessary in order to define affiliated density matrices and some form of generalized entropy. In particular is necessary in order to define any QES interpretation of the two factors as somehow associated with the entanglement wedges of the black hole and the radiation at each evaporation time $s$. Finally the transfer of information requires, at least intuitively, to understand how the evolution during the evaporation "moves" elements that initially are in $M_B(s_0)$, for some early time $s_0$, into elements in $M_R(s)$ for $s>s_0$ and larger than Page time \footnote{ Algebraically this transfer of information already imposes some conditions on how the evaporation process is implemented. Indeed imagine we denote $U_s$ the operator defining the evolution during the evaporation process in such a way that for some element $a\in M_B(s_0)$ for $M_B(s_0)$ representing the factor associated to the black hole entanglement wedge at the moment $s_0$ of black hole formation, we can define $a(s)=U_saU_s^{-1}$ as representing the operator $a$ at time $s$ after the formation. Transfer of information means that somehow for $s>s_P$, for $s_P$ representing Page time, the operator $a(s)$ is not in $M_B(s)$ but instead in $M_R(s)$. This means that $U_s$ cannot be an automorphism ( not even outer ) of $M_B(s_0)$.}.

Recently Engelhardt and Liu \cite{EL2} have suggested an interesting geometric model of the complexity transfer using non minimal QES's. The idea is to assume the appearance, at some early time, of some non minimal QES that effectively divides the black hole entanglement wedge into two pieces. Using Araki's basic result \cite{Araki} you can associate with both pieces two type $III_1$ factors and to represent the black hole entanglement wedge as {\it generated} by two type $III_1$ factors. The operators in one of these pieces are the candidates, on the basis of a {\it complexity} argument, to evolve in evaporation time into operators in the later radiation wedge algebra. In any case this construction requires, at least algebraically, to use some generalized entropy functional to define the non minimal QES and consequently requires some semifinite trace \footnote{Note that this is the case irrespectively what type $III_1$ factors we associate to the two pieces of the entanglement wedge.}. Apart from that the description in terms of a family of couples of hyperfinite type $III_1$ factors, and consequently all of them algebraically isomorphic, implies, as we will see, serious difficulties when identifying a {\it quantitative measure} of the {\it spatial} differences at different moments of the evaporation process.

\vspace{0.3 cm}

To start discussing this problem let us consider first the simplest possibility. 
 
 Using the notation of the previous section we could suggest as a basic model a one parameter family of couples of the type $II_{\infty}$ factors, associated to the hyperfinite type $III_1$ factor $\tilde{\cal{A}}$ using a {\it one parameter family of dominant weights $\omega_s$.} In other words we define the family of couples as
 \begin{equation}
 (\tilde{\cal{A}}_{\omega_s},\tilde{\cal{A}}_{\omega_s}^{'})
 \end{equation}
 An important, but necessary, restriction of this simple minded model is to require that $\omega_s$ are for all values of $s$ {\it dominant weights}\footnote{This is needed in order to identify the corresponding centralizers as type $II_{\infty}$ factors.}. This means that they are {\it equivalent} from the point of view of Connes-Takesaki equivalence of weights \cite{CT}. 
 
 The natural entropy functional $S(s)$ of our requirement ii) above, can be defined as the von Neumann entropy of an affiliated density matrix $\rho(s)$ defined by
 \begin{equation}
 \rho(s)=C(\omega_{s_0},\omega_s)
\end{equation}
For $C(\omega_{s_0},\omega_s)$ Connes cocycle relating the weights $\omega_{s_0}$ defined for some reference initial time $s_0$, and the weight $\omega_s$.

Let us now focus on how to identify {\it spatial differences} along the evaporation process when we use to represent $M_B(s)$ and $M_R(s)$, as we do in the previous simple model, hyperfinite type $II_{\infty}$ factors. In this case all the factors $M_B(s)$ as well as the factors $M_R(s)$ are {\it algebraically isomorphic to the hyperfinite type $II_{\infty}$ factor}. Thus we can associate to the black hole entanglement wedge and to the radiation entanglement wedge the algebraic classes $\bar M_B$ and $\bar M_R$. This information is not enough to characterize the spatial class of the couples $(M_B(s),M_R(s))$ at any evaporation time $s$. This spatial information requires to identify the Hilbert space on which these different factors are acting. The parameter $\theta_{MvN}$ defined in (\ref{deficit}) is the way, used in \cite{MvN4}, to characterize the spatial nature of $(M_B(s),M_R(s))$ at different evaporation times. If at any time the factors describing the black hole and the radiation are infinite factors i.e. type $II_{\infty}$ in the simplest example defined above or type $III_1$ factors in the case discussed in \cite{EL2} we find that the parameter $\theta_{MvN}$ has a formal value $\frac{\infty}{\infty}$ that is not any real number \footnote{See discussion in \cite{MvN4} in definition 3.3.1 and lemma 3.3.2.}. 

This difficulty, when we describe the black hole and the radiation using {\it infinite} factors, to identify quantitatively i.e. in terms of $\theta_{MvN}$, the spatial differences along the process of evaporation strongly suggest that the algebraic description of the black hole evaporation should be done using {\it finite} type $II_1$ factors. 

Next we will define a type $II_1$ model of black hole evaporation satisfying the Page curve and we will identify the transfer of information along the evaporation process in terms of Murray von Neumann parameter $\theta_{MvN}$.

\section{A type $II_1$ toy model of black hole evaporation}
We will associate with the evaporation process a one parameter family of couples $(M_B(s),M_R(s))$ of type $II_1$ factors with $M_R(s)=M_B(s)^{'}$. This is equivalent to a family of type $II_1$ factors $M_B(s)$ acting on a family of infinite dimensional Hilbert spaces $H_s$. Indeed, given $M_B(s)$ and $H_s$ the factor $M_R(s)$ is uniquely determined. For the time being we will denote the evaporation time by the formal parameter $s$. Its relation with physical time will be discussed later.

Associated to $M_B(s)$ and $H_s$ we define the Murray von Neumann coupling $dim_{M_B(s)}H_s$ that recall is given by $\frac{d_{M_B(s)}(M_R(s)|\psi\rangle)}{d_{M_R(s)}(M_B(s)|\psi\rangle)}$ for $|\psi\rangle$ {\it any} vector state in $H_s$. Recall that this quantity is independent of $|\psi\rangle$. We will denote this quantity $d_{MvN}(B,s)$. 

The importance of this quantity is that it encodes properties of the couple $(M_B(s),M_R(s))$ \footnote{In other words the crucial algebraic constraint defining the evaporation lies in requiring that at any moment of the process the factor associated to the black hole entanglement wedge is the commutant of the factor associated to the radiation entanglement wedge. In Murray von Neumann terminology \cite{MvN4} this defines a {\it spatial} property.} . The goal of this section will be to describe the evaporation process in terms of a function $d_{MvN}(B,s)$ defining the $s$ dependence of the coupling. The {\it dual} radiation description could be defined using $d_{MvN}(R,s)$. Recall that as discussed in section 2 we have at any $s$ 
\begin{equation}
d_{MvN}(B,s)=\frac{1}{d_{MvN}(R,s)}
\end{equation}
To define a concrete model we need to specify how these couplings depend on the evaporation time $s$. In this section we will simply present a {\it phenomenological model} where we assume that these couplings depend monotonically on $s$. In next section we will suggest the reason underlying this assumption. 

\subsection{Cyclic states}
An important property of the couplings $d_{MvN}(B,s)$ and $d_{MvN}(R,s)$ is the following:

\vspace{0.3 cm}

i) For times $s$ such that $d_{MvN}(B,s)<1$ ( or equivalently for $d_{MvN}(R,s)>1$)  it exists a vector state $|\Psi(s)\rangle \in H_s$ that is {\it cyclic} w.r.t. $M_{B}(s)$ and consequently {\it separating} w.r.t. $M_R(s)$.

\vspace{0.3 cm}

ii) For times $s$ such that $d_{MvN}(B,S)>1$ ( or equivalently for $d_{MvN}(R,s)<1$)  it exists a vector state $|\Psi(s)\rangle \in H_s$ that is {\it cyclic} w.r.t $M_{R}(s)$ and {\it separating} with respect $M_B(s)$.

\vspace{0.3 cm}

iii) For any time $s_P$ such that $d_{MvN}(B,s_P)=d_{MvN}(R,s_P)=1$ there exists a state $|\Psi(s_P)\rangle$ that is cyclic for both $M_{B}(s_P)$ and $M_{R}(s_P)$.

\vspace{0.3 cm}

Let us assume that at the initial time of black hole formation, let us say $s_0$, we have $d_{MvN}(B,s_0)<1$. This means that at this initial time we can find a pure state $|\Psi(s_0)\rangle$ that is {\it cyclic} w.r.t. the black hole entanglement wedge algebra $M_B(s_0)$. However at this initial time since $d_{MvN}(B,s_0)<1$ the state $|\Psi(s_0)\rangle$ is not {\it separating} w.r.t. $M_B(s_0)$. This means that there exist some operator $a_{B}\neq 0$ in $M_B(s_0)$ such that $a_{B}|\Psi(s_0)\rangle =0$. Reciprocally the state $|\Psi(s_0)\rangle$ is not cyclic w.r.t. the radiation algebra $M_R(s_0)$ but it is separating i.e. $a_R|\Psi(s_0)\rangle=0$ for $a_R\in M_R(s_0)$ implies $a_R=0$. Intuitively, although this is just a very imprecise picture, we can think of $|\Psi(s_0)\rangle$ as a sort of {\it $B$ vacuum} in the sense that is annihilated by some operators in $M_B(s_0)$.

Next assume ( recall that at this point we are just defining a purely phenomenological model ) that $d_{MvN}(B,s)$ {\it grows monotonically} with $s$ for $s>s_0$. Then and until we reach at $s_P$ the value $d_{MvN}(B,s_P)=1$ i.e. for $s<s_P$, we continue having a state $|\Psi(s)\rangle$ cyclic w.r.t. $M_B(s)$ but {\it not separating}. 

When we reach $s_P$ this state becomes cyclic w.r.t. both $M_B(s)$ and $M_R(s)$ i.e. the state $|\Psi(s_P)\rangle$ is at this time cyclic and {\it separating}. This is the point that we will identify as Page point. 

What happen when we move into times $s>s_P$ ? Since by assumption $d_{MvN}(B,s)$ grows monotonically we will find that after Page time $s_P$ we have $d_{MvN}(B,s)>1$ meaning that now we don't have a state cyclic w.r.t. $M_B(s)$. Instead we have a state that is {\it cyclic} w.r.t. the algebra associated to the radiation entanglement wedge i.e. $M_R(s)$. Note that in this {\it after Page's time phase} the state $|\Psi(s)\rangle$ is not separating w.r.t. $M_R(s)$ and therefore we can find non vanishing operators $a_R$ in $M_R(s)$ that annihilate the state $|\Psi(s)\rangle$. Using the analogy introduced above we can say that in this later phase we have an {\it $R$-vacuum}. 

Using these data we can define the {\it Page coupling} as
\begin{center}
\boxed{d^{P}_{MvN}(s) = \min\{d_{MvN}(B,s),d_{MvN}(R,s)\}}
\end{center}

that follows a curve of the type depicted in Figure 4. 

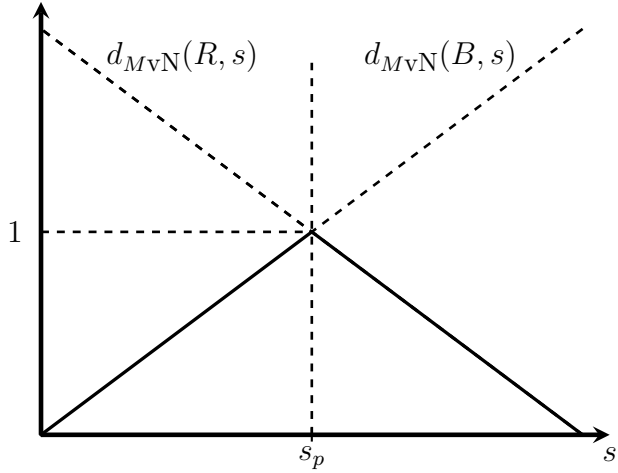
\begin{figure}
    \begin{center}
        \begin{tikzpicture}[scale=1.8]
        \draw[-stealth,thick,line width=1.75](0,0)--(4.2,0)node[anchor=north]{$s$};
        \draw[-stealth,thick,line width=1.75](0,0)--(0,3.2)node[anchor=east]{};
        \draw[thick,line width=1,dashed](4,3)--(2,1.5);
        \draw[thick,line width=1,dashed](4,0)--(0,3);
        \draw[thick,line width=1,dashed](0,1.5)--(2,1.5);
        \draw[thick,line width=1,dashed](2,0)--(2,2.75);
        \draw[thick,line width=1,dashed](4,0)--(0,3);
  \draw[thick]  (-0.05,1.5)circle(0pt) node[anchor=east ] {$1$};
        \draw[thick,line width=1.25](2,1.5)--(0,0);
        \draw[thick,line width=1.25](2,1.5)--(4,0);
        \filldraw(0,0) circle(.25pt);
        \draw[thick]  (4-.4,3-.4)circle(0pt) node[anchor=south east ] {$d_{M\mbox{\footnotesize vN}}(B,s)$};
        \draw[thick]  (.8-.4,3-.4)circle(0pt) node[anchor=south west ] {$d_{M\mbox{\footnotesize vN}}(R,s)$};
        \draw[thick]   (2,-0.05) --(2,0.05);
        \draw[thick]  (2,0)circle(0pt) node[anchor=north ] {$s_p$};
        \end{tikzpicture}
    \end{center}
    \caption{ Representation of Page coupling. $B$ phase corresponds to the region $[0,s_p]$ and $R$ phase to the region $s>s_p$.}\label{fig:4} 
\end{figure}

The intuitive way to understand this definition could be described as follows. At $s=s_P$ we perform a {\it duality} transformation $d_{MvN}(B,s_P)=\frac{1}{d_{MvN}(R,s_P)}$. For $s>s_P$ we describe the system in terms of {\it $R$-variables} i.e. in terms of $d_{MvN}(R,s)$. 

We can define the state $|\Psi^{cycl}(s)\rangle$ that for $s<s_P$ is cyclic with respect to $M_{B}(s)$ and for $s>s_P$
is cyclic for $M_{R}(s)$. Heuristically we can think of $|\Psi^{cycl}(s)\rangle$ as representing a {\it B-vacuum} for $s<s_P$ and a {\it R-vacuum} for $s>s_P$.
\subsection{Dual phases}
We will define two phases, namely {\it The $B$ Phase} defined by those times $s$ such that it exists a vector state cyclic w.r.t. $M_{B}(s)$ and the {\it The $R$ phase} defined by those times $s$ such that it exists a vector state cyclic w.r.t. $M_{R}(s)$. With these definitions the time $s_P$ defines a change of phase. 

\subsection{The algebraic Page curve}
Using the previous type $II_1$ {\it model of evaporation} we can define a Page curve using the relative dimensions as follows:

\vspace{0.3 cm}

i) At initial time $s_0$ where we require $d_{MvN}(B,s_0)<1$ we start in {\it $B$ phase} with a vector state $|\Psi(s_0)\rangle$ cyclic w.r.t. $M_{B}(s_0)$. At this point we have for the relative dimensions the relation 
\begin{equation}\label{pagerelation}
\log (d_{M_{B}(s_0)}(M_{R}(s_0) |\Psi(s_0)\rangle) = \log (d_{M_{R}(s_0)}(M_{B}(s_0) |\Psi(s_0)\rangle) + \log d_{MvN}(B,s_0)
\end{equation}

\vspace{0.3 cm}

ii) For $s>s_0$ but smaller than $s_P$ we have assumed $d_{MvN}(B,s)<1$. In other words we remain in the {\it $B$ phase} and we have a state $|\Psi(s)\rangle$ cyclic w.r.t. $M_{B}(s)$.

\vspace{0.3 cm}

iii) For $s=s_P$ we have a vector $|\Psi(s_0)\rangle$ that is cyclic for both $M_{B}(s_P)$ and $M_{R}(s_P)$. This will represent the Page point.

\vspace{0.3 cm}

iv) For $s>s_P$ we enter into the {\it $R$ phase} i.e.
 $d_{MvN}(R,s) <1$ and consequently there exists a state $|\Psi(s)\rangle$ cyclic w.r.t. $M_{R}(s)$. In this phase we get again (\ref{pagerelation}). The only difference between both phases is that while in the $B$ phase we have that $d_{M_B(s)}(M_R(s)|\Psi(s)\rangle) < d_{M_R(s)}(M_B(s)|\Psi(s)\rangle)$ the situation is the {\it reverse} in the $R$ phase.

\vspace{0.3 cm}

v) Finally we can define a type $II_1$ Page curve by
\begin{equation}
\min(\log (d_{M_{B}(s)}(M_{R}(s) |\Psi^{cycl}(s))\rangle),\log (d_{M_{R}(s)}(M_{B}(s) |\Psi^{cycl}(s)\rangle))
\end{equation}

In summary in this algebraic version of the Page curve what we describe is the evolution of the relative (black hole versus radiation) continuous dimension along the evaporation process. The phenomenological input we introduce is the monotonous growth of the coupling \footnote{For a preliminary version of this model see \cite{gomez1}}.

\subsection{An algebraic representation of QES: transfer of information}
Let us now focus on the {\it spatial} differences, in this toy model, along the evaporation process. These differences are, as discussed, determined by the behaviour of the parameter $\theta_{MvN}$ along the evaporation process. Now since we are working with type $II_1$ factors this parameter can take any value in $[0,\infty]$. The limit cases $\theta_{MvN}=0$ and $\theta_{MvN}=\infty$ correspond to the cases where either $M_B$ or $M_R$ are type $II_{\infty}$ factors.

 Next we will identify the transfer of {\it "information"} in terms of $\theta_{MvN}(s)$.

In order to do that let us define an information transfer parameter $t_{inf}(s)$ as
\begin{equation}\label{definf}
t_{inf}(s)=\frac{d_{M_R(s)}(M_B(s)|\Psi(s)\rangle)}{d_{M_R(s)}(H_s)}
\end{equation}
i.e. as the ratio, at time $s$, between the size of the black hole entanglement wedge $(M_B(s)|\Psi(s)\rangle)$ {\it measured} by the radiation i.e. $d_{M_R(s)}(M_B(s)|\Psi(s)\rangle)$ and the size of the full Hilbert space $H_s$ equally measured by the radiation i.e. $d_{M_R(s)}(H_s)$.

In the $B$ phase i.e. for $s<s_P$ since we can choose $|\Psi(s)\rangle$  cyclic w.r.t. $M_B(s)$ i.e. $M_B(s)|\Psi(s)\rangle =H_s$\footnote{In the sense that $M_B(s)|\Psi(s)\rangle$ is dense in $H_s$.} we get
\begin{equation}
t_{inf}(s)=1
\end{equation}
In the $R$ phase we instead get
\begin{equation}\label{info}
t_{inf}(s) = \frac{1}{\theta_{MvN}(s)}
\end{equation}
with $\theta_{MvN}(s)$ defined in (\ref{deficit}) i.e. in present notation $\theta_{MvN}(s)= d_{MvN}(B,s)\frac{d_{M_R(s)}(H_s)}{d_{M_B(s)}(H_s)}$. 
To derive (\ref{info}) we simply rewrite (\ref{definf}) as
\begin{equation}
t_{inf}(s)=\frac{d_{M_B(s)}(M_R(s)|\Psi(s)\rangle)}{d_{MvN}(B,s) d_{M_R(s)}(H_s)}
\end{equation}
Using now that in the $R$ phase $|\Psi(s)\rangle$ is {\it cyclic} w.r.t. $M_R(s)$ we get
\begin{equation}
t_{inf}(s)=\frac{d_{M_B(s)}(H_s)}{d_{MvN}(B,s) d_{M_R(s)}(H_s)}
\end{equation}
which gives rise to (\ref{info}).

Note that as advertised the transfer of information along the evaporation process has a well defined value only when we work with finite type $II_1$ factors. For the eternal black hole defined in terms of two type $II_{\infty}$ factors this transfer of information becomes the entity $\frac{\infty}{\infty}$ different from any real number ( see comment in \cite{MvN4} after def 3.3.1).

We can now define $T_{inf}(s)=-\log t_{inf}(s)$ that is $0$ in the $B$ phase and $\log \theta(s)$ in the $R$ phase. 

Hence we can use $T_{inf}(s)$ as an {\it order parameter} of the Page phase transition (see figure 5). 
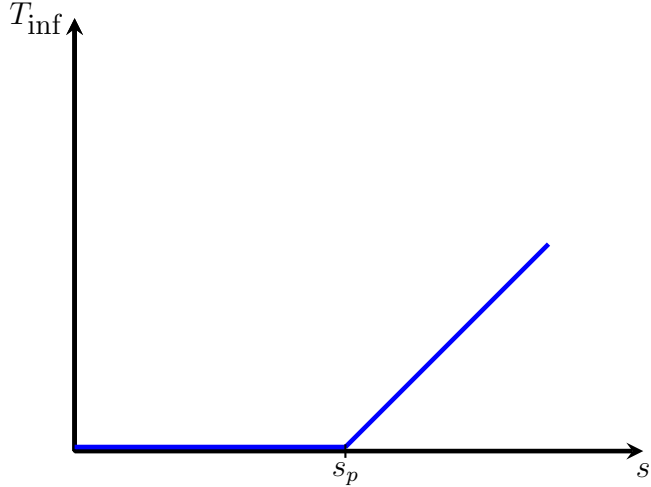
\begin{figure}
    \begin{center}
        \begin{tikzpicture}[scale=1.80]
            \draw[-stealth,thick,line width=1.75](0,0)--(0,3.2)node[anchor=east]{$T_{\mbox{\small inf}}$};
        \draw[-stealth,thick,line width=1.75](0,0)--(4.2,0)node[anchor=north]{$s$};
        \draw[-stealth,thick,line width=1.75](0,0)--(0,3.2)node[anchor=east]{};
        \draw[thick]  (2,0)circle(0pt) node[anchor=north ] {$s_p$};
        \draw[thick,line width=1.75,blue](0,0.03)--(2,0.03);
        \draw[thick,line width=1.75,blue](2,0.03)--(3.5,1.53);
        \draw[thick]   (2,-0.05) --(2,0.05);

        \filldraw(0,0) circle(.25pt);

        \end{tikzpicture}
    \end{center}
    \caption{Order parameter for Page phase transition}\label{fig:5} 
\end{figure}

The former discussion leads to a natural ( formal) relation with the QES model of information transfer of the type
\begin{center}
\boxed{\frac{Area(\chi(s))}{4G_N} \sim T_{inf}(s) = \log \theta_{MvN}(s)}
\end{center}
that could be interpret as the {\it algebraic representation of QES}.

{\bf Remark} In \cite{EF} ( see also \cite{EL2} ) the question about the differences between two times $s_1<s_P<s_2$ sharing the same value of the generalized entropy arises. Note that the form of the Page curve always implies the existence of those times. In \cite{EL2} this question is discussed from the point of view of the EPR$=$ER conjecture\cite{EPR} and the emergency of space time connectivity. In the present scheme the difference between these times is determined by the value of $T_{inf}$ that is zero for $s_1$ and different from zero for $s_2$.

\section{The fundamental group and the black hole evaporation}
In the previous phenomenological model we have left unanswered several important questions, namely

{\bf Q-1} How the type $II_1$ factors $M_B(s)$ and $M_R(s)$ associated to the entanglement wedges of the black hole and the radiation are defined.

{\bf Q-2} What is the dynamics underlying the $s$ dependence of these $II_1$ factors.
 
{\bf Q-3} What is the physical meaning of the evolution parameter $s$.

In this section we will address some of these questions. 

Concerning Q-1 we will identify the couple $(M_B(s),M_R(s))$ of type $II_1$ factors as
\begin{equation}\label{1factor}
M_B(s)=P_s{\cal{N}}P_s
\end{equation}
for ${\cal{N}}$ the hyperfinite type $II_{\infty}$ factor acting on a Hilbert space $\hat H$ and with $P_s$ finite projections in ${\cal{N}}$. Denoting $\tau_{\cal{N}}$ the semifinite trace on ${\cal{N}}$ these projections are characterized by $\tau_{\cal{N}}(P_s)=s$ with $s\in[0,\infty]$. The factor $M_B(s)$ is acting on $P_s\hat H$. Relative to $\hat H$ we define $M_R(s)$ as the commutant of $M_B(s)$. This defines the couple $(M_B(s),M_R(s))$.

The factor ${\cal{N}}$ can be defined as follows. To each entanglement wedge $W_B(s)$ and $W_R(s)$ you associate two type $III_1$ factors let us say ${\cal{A}}_B(s)$ and ${\cal{A}}_R(s)$ which are {\it algebraically isomorphic among them}\footnote{This follows from the algebraic uniqueness of the hyperfinite type $III_1$ factor.}. Now you can define the isomorphic couples $(\tilde {\cal{A}}_B(s),\tilde {\cal{A}}_R(s))$ with $\tilde{\cal{A}}={\cal{A}}\otimes F_{\infty}$ and define ${\cal{N}}$ as the corresponding centralizer for a dominant weight $\omega=\phi\otimes \bar\omega$ for $\phi$ a state on $\cal{A}$. This construction identifies $\tau_{\cal{N}}$ with the trace induced by $\omega$ on the centralizer ${\cal{A}}_{\omega}$. The Hilbert space $\hat H$ on which the so defined $\cal{N}$ is acting is $\hat H=L^2(\mathbb{R},H_{\phi})$ for $H_{\phi}$ the GNS Hilbert space defined by the state $\phi$ used in the definition of the dominant weight $\omega$. This set of steps defines the factor $\cal{N}$ used in (\ref{1factor}). 

The obvious question at this point of the construction is how all this depends on the particular state $\phi$ used to define $\omega$. As we will discuss in the next section the natural physics interpretation of the GNS Hilbert space $H_{\phi}$ on which the type $III_1$ factor ${\cal{A}}_B(s)$ is acting is to represent the {\it Fock like} space of $O(1)$ \footnote{Including $G_N$ we can interpret these fluctuations as $O(G_N^{0})$.} quantum fluctuations,  around a given background defined by $\phi$, and  localized on the wedge $W_B(s)$. Different choices $\psi$ of the state defining the dominant weight can represent effects of order $O(\frac{1}{G_N})$ with respect to $\phi$. To identify these effects we need to define a map between the Hilbert spaces $H_{\phi}$ and $H_{\psi}$ of $O(1)$ quantum fluctuations around $\phi$ and $\psi$. Since ${\cal{N}}$ depends on the state $\phi$ as a crossed product, namely ${\cal{N}}={\cal{A}}\rtimes_{\sigma_{\phi}}{\mathbb{R}}$, changes of $\phi$ are naturally implemented unitarily in $\hat H$ using the cocycle between $\phi$ and $\psi$. In other words different "backgrounds" define {\it spatially isomorphic} type $II_{\infty}$ factors ${\cal{N}}$. 

But, What this means physically ? It means that although we can use different states $\phi_s$ to describe the $O(1)$ quantum fluctuations on the entanglement wedges at different times of the evaporation process, the corresponding couple of type $II_{\infty}$ factors $({\cal{N}}_B(s),{\cal{N}}_R(s))$ cannot be distinguished {\it spatially}. Recall that in this case the parameter $\theta_{MvN}$ is unique and fixed to be the  entity $\frac{\infty}{\infty}$. This explains why black hole evaporation, by contrast to eternal black holes, cannot be described using {\it infinite factors}. 

After these comments let us come back to the finite factors $M_B(s)$ and $M_R(s)$ defined by (\ref{1factor}). This leads to our question Q-2. In order to describe black hole evaporation we need to identify a family of couples $(M_B(s),M_R(s)$ representing spatial differences for different evaporation times. Recall that these spatial differences are the ones we describe using the Page curve and also the ones underlying the transfer of information along the process.

As already discussed in the toy model presented in previous section we can use, in order to define the spatial properties, the Murray von Neumann couplings $d_{MvN}(B,s)$ or its dual $d_{MvN}(R,s)$. Note that these couplings are now defining {\it intrinsic} properties of the couples $(M_B(s),M_R(s))$ i.e. independent on normalization conventions. Moreover these couplings depend on the trace $\tau_{\cal{N}}$.

In order to model the $s$ dependence of these finite factors let us consider the Takesaki automorphisms $\theta_{\lambda}$ of $\cal{N}$ scaling the trace $\tau_{\cal{N}}$ by $\lambda$. The action of these automorphisms on the finite projections $P_s$ is determined by
\begin{equation}
\tau_{\cal{N}}(\theta_{\lambda}(P_s))=\lambda s
\end{equation} 
and therefore they change the factor $M_B(s)$ defined in (\ref{1factor}) as
\begin{equation}\label{transformation}
M_B(s)\rightarrow M_B(\lambda s)
\end{equation}
If we are considering for $\cal{N}$ the hyperfinite type $II_{\infty}$ factor the transformation defined in (\ref{transformation}) defines an element of the fundamental group $\cal{G}$ \cite{Takesaki} of the associated type $II_1$ factor. In the present case \footnote{Since we are assuming hyperfiniteness.} the fundamental group \cite{MvN4} is just $\mathbb{R}^{+}$ i.e. $[0,\infty]$.

Now we can come back to our type $II_1$ model of the Page curve described in the previous section. Thus we can start at some initial time $s_0$ associated with the black hole formation in the $B$-phase i.e. with $d_{MvN}(B,s_0)<1$. Acting with an element $\lambda \in {\cal{G}}$ we get the factor $M_B(\lambda s_0)$. Let us define an evaporation time $\hat s$ as
\begin{equation}
\lambda=e^{-\hat s}
\end{equation}
with $\hat s\in [0,\infty]$. So, relative to the initial condition at $s_0$, the factor at "time" $\hat s$ is given by
\begin{equation}
M_B(\hat s)= \theta_{\lambda}(P_{s_0}) {\cal{N}}\theta_{\lambda} (P_{s_0})
\end{equation}
for $\lambda=e^{-\hat s}$. In these conditions we get, using (\ref{projection}), that
$d_{MvN}(B,\hat s)= d_{MvN}(B,s_0)e^{\hat s}$. Thus the time $\hat s\in [0,\infty]$ ( measuring the time after black hole formation) is associated with the action of the element in the fundamental group $\cal{G}$ equal to 
$e^{-\hat s}$. 

The Page time ${\hat s}_P$ is, following our model in section 5, defined by the condition 
\begin{center}\label{pagetime}
\boxed{e^{{\hat s}_P} d_{MvN}(B,s_0)=1}
\end{center}
Note that in this type $II_1$ version of black hole evaporation two algebraic facts are crucial, namely the finitude of the factors representing the entanglement wedges of the black hole and the radiation and the non triviality of the corresponding {\it fundamental group} \footnote{ Note that the Page time defined by (\ref{pagetime}) depends on the initial condition $d_{MvN}(B,s_0)$. Qualitatively we can define this initial value as $e^{-N_{s_0}}$ for $N_{s_0}$ the average, after a Planck time, of quanta of "wave length" $O(R)$, for $R$ the black hole size at the moment of formation. Using very roughly the Stefan- Boltzmann-Hawking law we get $N_{s_0}\sim \sqrt{N_{BH}}$ for $N_{BH}\sim \frac{R^2}{L_P^2}$. In this case $s_P$ defined by (\ref{pagetime}) would scales like $\sqrt{N_{BH}}$.}. 

Finally we need to identify the meaning of $\hat s$. If we use the Weyl generator $V_{\hat s}$ canonically conjugated to $U_t$ to define the automorphism $\theta_{\hat s}$ we can identify $\hat s$ as the {\it dual} to the "time" parameter $t$ in $U_t$. Since this $t$ can be identified with the modular time defining the modular automorphism $\sigma_{\omega}^t$ we can think the evaporation time $\hat s$ as the dual to the modular time $t$.
\section{A speculative comment on the algebraic meaning of $G_N$}
Generically and after coupling to gravity we can define the factor $M$ representing the diff invariant gravitationally dressed operators \footnote{For instance we can define in the notation used in this paper $M=E_{\phi}{\tilde{\cal{A}}}_{\omega}$ for $E_{\phi}$ the conditional expectation defined by the crossed product representation (\ref{crossedw}) of ${\tilde{\cal{A}}}_{\omega}$.}. For a given state $\phi$ on $M$ we can define, as the Hilbert space on which $M$ is acting the corresponding GNS Hilbert space $H_{\phi}$. This Hilbert space represents $O(1)$ diff invariant quantum fluctuations on the background defined by the state $\phi$ or equivalently it describes the Hilbert space in the limit $G_N=0$. To describe standard perturbative effects scaling like $O(G_N^a)$ with $a>0$ i.e. those effects that go to zero in the $G_N=0$ limit, as can be the standard perturbative interaction between gravitons, can be in principle done using standard techniques in terms of perturbative series in positive powers of $G_N$ \footnote{Note that these perturbative interactions are not described by the algebraic modular dynamics.}

In quantum gravity we are interested in defining non perturbative $O(\frac{1}{G_N})$ effects. The algebraic version of these effects in the limit $G_N=0$ appears as vanishing quantum superpositions of states in different GNS Hilbert spaces $H_{\phi}$ when the states $\phi$ represent backgrounds perturbatively unrelated.

The question we want to address in this final comment is how to quantify {\it algebraically} what they would be $O(\frac{1}{G_N})$ non perturbative effects. We will suggest a very speculative answer based on the {\it spatial} properties of finite type $II_1$ factors.

In a nutshell the idea is to define an {\it algebraic} dimensionless analog $\tilde G_N$ of $G_N$ measuring the {\it spatial} properties of the type $II_1$ factor $M$ defining the gravitationally dressed operators. We will simply suggest the formal definition
\begin{equation}\label{crazy}
\tilde G_N\sim \frac{1}{\theta_{MvN}}
\end{equation}
The reason underlying the former identification is to associate the formal limit $\tilde G_N=0$ with $\theta =\infty$ that is what we expect for the case where $M$ is finite and the commutant $M^{'}$ is infinite ( see \cite{MvN4} lemma 3.3.2 ($\gamma$).

The identification (\ref{crazy}) is close in spirit to the suggestion in \cite{Witten3} on the connection of $G_N$ and the Murray von Neumann coupling for the case of dS. If now we use the relation between $\theta_{MvN}$, in the black hole evaporation process, with the area of the QES, suggested in the previous section we get
\begin{equation}
\tilde G_N\sim e^{-\frac{Area(\chi_s)}{4G_N}}
\end{equation}
with $\tilde G_N$ depending on the evaporation time $s$. As stressed this is just a speculative comment that should be taken with a grain of salt.

\section{Acknowledgments}
I would like to thank Hong Liu for useful discussions.
This work was supported by grants SEV-2016-0597, FPA2015-65480-P and PGC2018-095976-B-C21.

\end{document}